**Thermal Transport in 3D Nanostructures**

*Haifei Zhan, Yihan Nie, Yongnan Chen\*, John M. Bell, and Yuantong Gu\**

Dr. H.F. Zhan, Mr. Y.H. Nie, Prof. J.M. Bell, Prof. Y.T. Gu
School of Chemistry, Physics and Mechanical Engineering, Queensland University of Technology, Brisbane QLD 4001, Australia
E-mail: yuantong.gu@qut.edu.au

Prof. Y.N. Chen
School of Materials science and Engineering, Chang'an University, Xi'an 710064, China
E-mail: frank_cyn@163.com

This work summarizes recent progress on the thermal transport properties of three-dimensional (3D) nanostructures, with an emphasis on experimental results. Depending on the applications, different 3D nanostructures can be prepared or designed to either achieve a low thermal conductivity for thermal insulation or thermoelectric devices, or a high thermal conductivity for thermal interface materials used in the continuing miniaturization of electronics. A broad range of 3D nanostructures have been discussed, ranging from colloidal crystals/assemblies, array structures, holey structures, hierarchical structures, 3D nanostructured fillers for metal matrix composites and polymer composites. Different factors that impact the thermal conductivity of these 3D structures are compared and analyzed. This work provides an overall understanding of the thermal transport properties of various 3D nanostructures, which will shed light on the thermal management at nanoscale.

## 1. Introduction

Thermal transport is one of the fundamental characteristics that determine the applications of materials. Depending on the application, materials are required to have either a high thermal conductivity, or a strongly suppressed thermal conductivity.[1] For energy saving in both residential and commercial buildings, there has been a continuing search for high-performance, light weight, and mechanically strong, thermally insulating materials, where a low thermal conductivity is required (**Figure 1**a).[2] Good thermal insulation is also required for the electrical, optical, and space applications in order to tightly regulate heat transfer during operation. Another major application that can be affected dramatically by the thermal transport properties is the direct energy conversion from heat to electricity using thermoelectric materials,[3] such as bismuth telluride based-electronics for low temperature applications.[4] The performance of the thermoelectric devices is measured by the figure of merit (*ZT*) as calculated from $ZT = S^2 \kappa_e \kappa_t^{-1} T$, where $S$ is the Seebeck coefficient; $\kappa_e$ and $\kappa_t$ are the electrical and thermal conductivity, respectively; and $T$ is the average temperature within the material. To obtain a high *ZT* with optimum efficiency, there are generally two strategies, including phonon engineering and band engineering. The purpose of phonon engineering is to reduce lattice thermal conductivity without degrading the electrical properties.[3b] Thermoelectric devices possess a vast potential for power generation from solar, automobile, industrial heat sources, and even human body heat (Figure 1b).[5]

On the other hand, the continuing miniaturization of electronic devices/systems (such as light-emitting diodes, integrated circuits and microprocessors) has inevitably increased their power density ($\sim 100$ Wcm$^2$) and led to dramatically increased heat generation.[6] To ensure reliable performance and lifetime, effective and efficient heat removal is desired, which relies heavily on the high thermal conductivity of the packaging substrates and thermal interface materials (Figure 1c).[7] The thermal interface materials are normally based on polymers, whose intrinsic thermal conductivity is only about 0.2-0.5 Wm$^{-1}$K$^{-1}$, being restricted by the strong inherent phonon scattering between chain ends, entanglements and impurities.

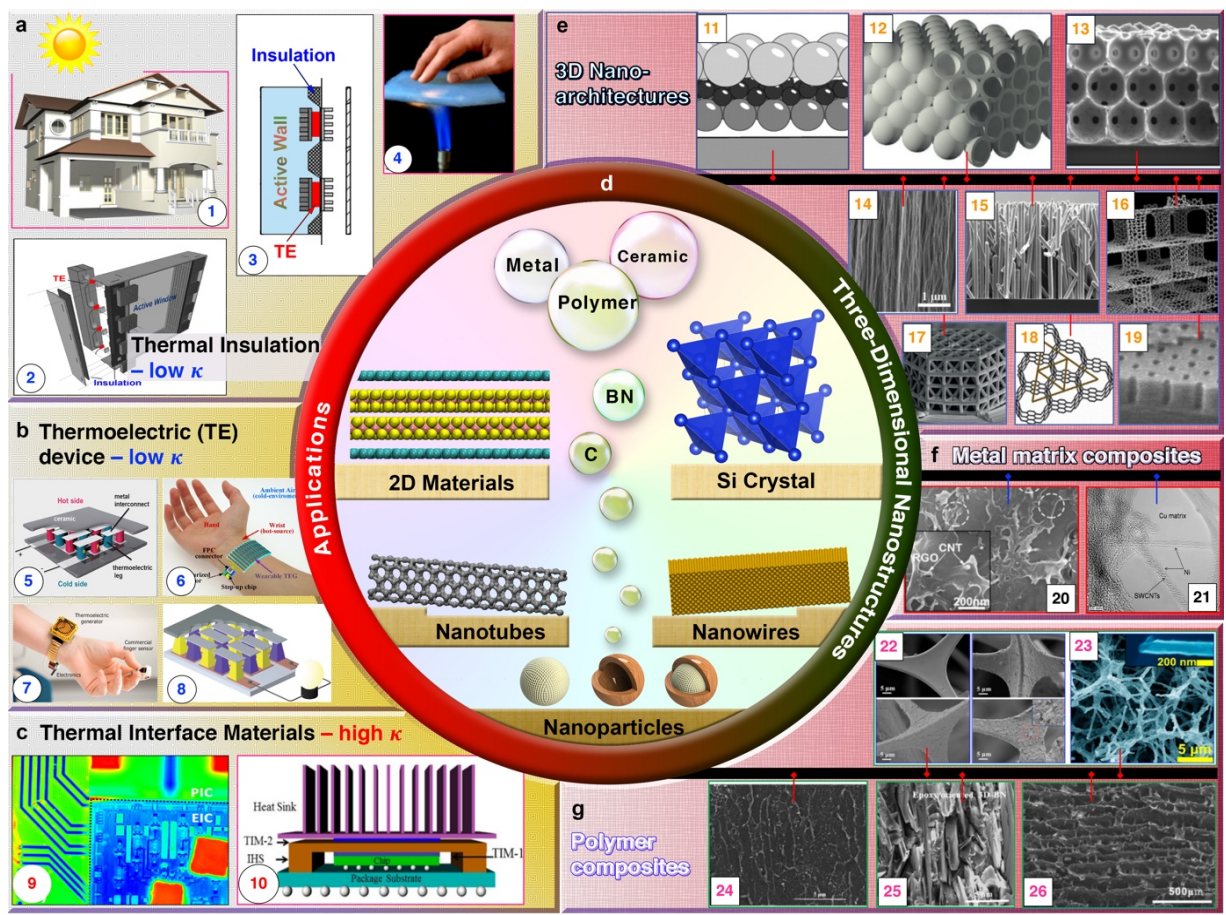

**Figure 1.** The palette of 3D nanostructures for thermal management. (a) Example applications for thermal insulation, including the house (1 – Ref.[8]), the active window (2 – Ref.[9]), the active wall (3 – Ref.[9]), and the aerogel above a blow torch (4 – Ref.[10]). (b) Example application of thermoelectric devices, including a thermoelectric module sketch (5 – Ref.[11]), a wearable thermoelectric generator (6 – Ref.[12]), a body-powered wireless pulse oximeter (7 – Ref.[13]), a thermoelectric generator powered flat bulk (8 – Ref.[14]). (c) Electronics, including a printed circuit board (9 – Ref.[15]), a schematic view of the electronic packing (10 – Ref. [16]). (d) Different low-dimensional nanomaterials. (e) Different 3D nanoarchitectures, including colloidal assemblies (11 – Ref.[17]), silica hollow sphere colloidal crystals (12 - Ref.[18]), Cu inverse opals (13 – Ref.[19]), vertically aligned carbon nanotube array (14 – Ref.[20]), aligned Cu nanowire arrays (15 – Ref.[21]), the atomic model of a pillared graphene (16 – Ref.[22]), a carbon nanolattice (17 – Ref.[23]), the atomic model of a carbon honeycomb

structure (18 – Ref.[24]), and a holey silicon (19 – Ref. [25]). (f) Different metal matrix composites, including Cu matrix with carbon nanotube and reduced graphene oxide network (20 – Ref.[26]), Cu matrix with embedded carbon nanotubes (21 -Ref.[27]). (g) Polymer composites with 3D nanostructures, including melamine foam with BN nanosheets assemblies (22 – Ref.[28]), 3D BN cellular architecture (23 – Ref.[29]), polystyrene composite with 3D segregated double networks (24 – Ref.[30]), epoxy composite with 3D BN network (25 – Ref.[31]), and 3D-BN nanosheets aerogels (26 – Ref.[32]). 1- reproduced with permission.[8] Copyright (2004), Elsevier. 2 and 3- reproduced with permission.[9] Copyright (2015), Elsevier. 4- reproduced with permission.[10] Copyright (2017), Elsevier. 5- reproduced with permission.[11] Copyright (2012), RSC publishing. 6- reproduced with permission.[12] Copyright (2018), Elsevier. 7- reproduced with permission.[13] Copyright (2009), AIP publishing. 8- reproduced with permission.[14] Copyright (2017), Wiley. 9- reproduced from reference.[15] 10- reproduced with permission.[16] Copyright (2012), Elsevier. 11- reproduced with permission.[17] Copyright (2015), American Chemical Society. 12- reproduced with permission.[18] Copyright (2017), Wiley. 13- reproduced with permission.[19] Copyright (2016), American Chemical Society. 14- reproduced from reference.[20] 15- reproduced with permission.[21] Copyright (2015), American Chemical Society. 16- reproduced with permission.[22] Copyright (2008), American Chemical Society. 17- reproduced with permission.[23] Copyright (2016), Springer Nature. 18- reproduced with permission.[24] Copyright (2016), American Physical Society. 19- reproduced with permission. [25] Copyright (2016), American Chemical Society. 20- reproduced with permission.[26] Copyright (2019), Elsevier. 21- reproduced with permission.[27] Copyright (2016), Elsevier. 22- reproduced with permission.[28] Copyright (2018), Elsevier. 23- reproduced with permission.[29] Copyright (2017), American Chemical Society. 24- reproduced with permission.[30] Copyright (2017),

American Chemical Society. 25- reproduced with permission.[31] Copyright (2017), American Chemical Society. 26- reproduced with permission.[32] Copyright (2015), Wiley.

To facilitate various thermal management requirements, significant research efforts have been devoted to either screen the materials with desired thermal transport properties or design novel materials/structures that satisfy different requirements.[33] The advance of nanotechnology enables the construction of novel materials from the bottom up, based on low-dimensional nanomaterials (Figure 1d), including zero-dimensional (e.g., fullerene and nanoparticles), one-dimensional (e.g., nanotubes, nanowires, and nanothreads),[34] and two-dimensional (e.g., nanoribbons or nanosheets)[35] nanomaterials. This work brings together an overview of the recent progress on the thermal transport properties of different three-dimension (3D) nanostructures with a focus on experimental studies, ranging from 3D nanoarchitectures (Figure 1e), metal-matrix composites with nanostructured fillers (Figure 1f), and the polymer composites with nanostructured fillers (Figure 1g). The discussions are conceptually divided into three major parts, i.e., 3D nanostructures for low thermal conductivity, 3D nanostructures for high thermal conductivity, and polymer composites with nanostructured fillers for high thermal conductivity. Sections in each part are organized based on the complexity of the nanostructures. The preparation or fabrication method of the corresponding nanostructures will be briefly introduced before discussing their thermal transport properties.

## 2. Three-dimensional Nanostructures for Low Thermal Conductivity

There is a plethora of 3D nanostructures being synthesized or theoretically predicted to meet low thermal conductivity requirements, which are normally constructed from low-dimensional nanomaterials (e.g., 0D fullerene, 1D nanowire or nanotube, and 2D nanosheet or nanoribbon). In subsequent sections, the 3D nanostructures with low thermal conductivity are critically discussed and generally divided into two categories considering their applications, i.e., for thermal isolation and thermoelectric devices.

## 2.1. Low Thermal Conductivity for Thermal Isolation

Thermal insulating materials have been widely used in our daily lives, such as insulating gloves, pipes and buildings.[36] To facilitate various advanced thermal insulating usages, several different types of 3D nanostructures with a low thermal conductivity have been synthesized or prepared. Based on their constituent components or building blocks, these 3D nanostructures can be categorized as: colloidal crystals and assemblies based on 0D nanostructures; 3D hierarchical structures/networks constructed from a combination of low-dimensional nanomaterials; and highly porous nanostructures.

### 2.1.1 Polymer Colloidal Crystals and Assemblies

Colloidal crystals are typically prepared through the bottom-up approach with a highly ordered hierarchical structures of colloidal particles.[17] They can be assembled into 2D films and 3D structures with either a single type of particles or multiple types of particles, and close-packed or non-close-packed with controlled defects. Different techniques have been developed to fabricate 3D colloidal crystals, such as sedimentation,[37] electrodeposition,[38] shear alignment,[39] and filtration.[40] Their hierarchical nature endows several unique features, including wide variability of material composition, scalability, tunable symmetry (amorphous or ordered crystals), and a large number of interfaces.[41] Extensive research efforts have been devoted to colloidal crystals over the past 30 years, covering the constituent materials, particle-particle interactions, applications (e.g., sensors, waveguides), and physical properties (e.g., the opalescent color) of the colloidal crystals.[17, 42] However, their thermal transport properties have rarely been investigated.

**Figure 2**a shows the types of structural hierarchies with colloidal particles in two- and three-dimensions. Generally, monodispersed spheres favor close-packed face-center-cubic (fcc) lattice geometry with particle volume fraction of 74% and 26% of interstitial space. The large number of interfaces always induce strong interface phonon scattering and lead to low thermal conductivity ($\kappa$). This feature makes them ideal for thermal insulating applications. For instance, a glass substrate with a hollow silicate particles coating has been reported to be a thermal insulator, with a reduction of around 30% in $\kappa$ compared to the uncoated substrate.[43] Colloidal particles can be synthesized as a solid sphere, core-shell sphere, or a hollow sphere with different compositions. Polystyrene (PS) and poly(methyl methacrylate-co-n-butyl acrylate) (PMMA) are the common materials for solid particle colloidal crystals (Figure 2b), and the silica colloidal crystals are normally constructed from core-shell particles. The thermal conductivity of colloidal crystals is calculated from $\kappa = \alpha C_p \rho$, where $C_p$ and $\rho$ are the heat capacity and density; $\alpha$ is the thermal diffusivity as determined from laser flash analysis. In general, the polymer solid particle-based colloidal crystals have been reported with low thermal conductivity, e.g., $\sim 51 \pm 6$ mWm$^{-1}$K$^{-1}$ for a PS colloidal crystal at 25 °C,[44] and $84 \pm 2$ mWm$^{-1}$K$^{-1}$ for PMMA (20vol% $n$-butyl acrylate ($n$BA)) colloidal crystal at 25 °C.[45]

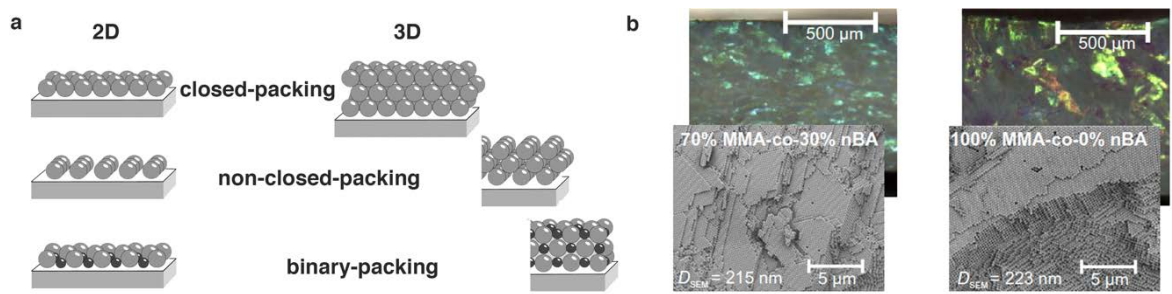

**Figure 2.** The colloidal crystals/assemblies. (a) Structural hierarchies of colloidal particles in two- and three-dimensions. Reproduced with permission.[17] Copyright 2015, American Chemical Society. (b) Optical and Scanning Electron Microscope (SEM) images of PMMA colloidal crystals. Reproduced from Ref.[46]

***Temperature dependence:*** Due to the softening of polymer particles above their glass transition temperature ($T_g$), the thermal conductivity of the colloidal crystals has a unique temperature dependence. **Figure 3**a compares the thermal conductivity of PS colloidal crystals with an initial thickness of 1094 μm.[44] The initial particulate system transfers to a continuous film when the temperature exceeds $T_g$ (around 105 °C), and a significant thickness drop around 84%-89% is observed. As illustrated in Figure 3a (during the first measuring cycle with increasing temperature from 25 °C), the thermal conductivity of the native colloidal crystals, which is around 50 mWm$^{-1}$K$^{-1}$, has a weak relationship with the temperature below $T_g$. Due to the sudden thickness drop above $T_g$, the thermal conductivity increases to around 150 mWm$^{-1}$K$^{-1}$, which is retained in the following cooling cycle from $T_g$ to 25 °C. Similar results are observed from the PMMA colloidal crystals as illustrated in Figure 3b,[47] the thermal conductivity experiences a step-like, irreversible increase from around 120 mWm$^{-1}$K$^{-1}$ to ~ 200 mWm$^{-1}$K$^{-1}$ at the transition temperature (around 74 °C). It is observed that the initial contact points between colloidal particles are enlarged after reaching the transition temperature (Figure 3c), which reduces the thermal interfacial resistance and leads to a remarkable increase to their thermal conductivity, e.g., up to ~ 200% increase.

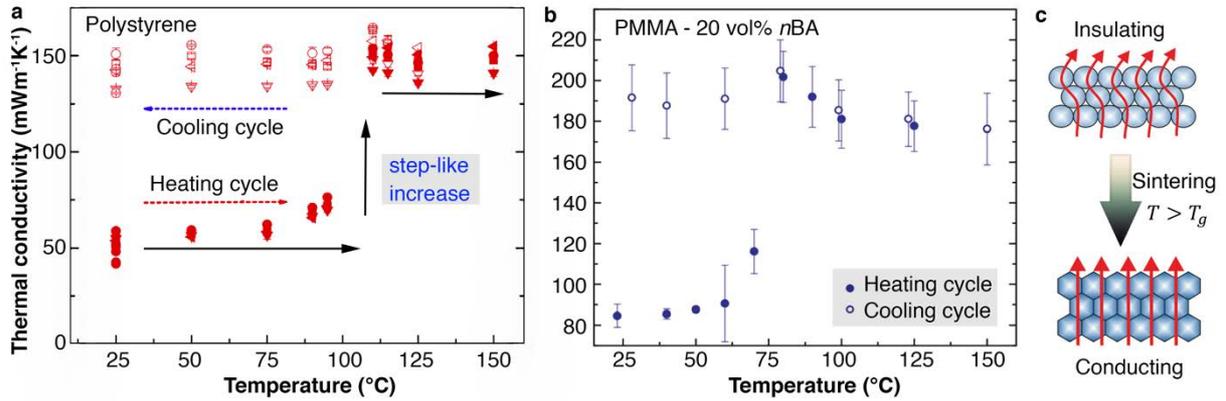

**Figure 3.** Thermal transport in polymer colloidal crystals. (a) The thermal conductivity of PS colloidal crystals as a function temperature. The particle has a diameter of 366 nm. The closed and open markers denote the heating and the cooling cycles, respectively. Reproduced with permission.[44] Copyright (2015), Elsevier. (b) The thermal conductivity of PMMA colloidal crystals with 20vo% of *n*BA. The particle has a diameter of $214 \pm 7$ nm. (c) Schematic illustration of the change of interfacial contacts between polymer colloidal crystals after reaching glass transition temperature. (b) and (c) are reproduced from Ref.[47]

The intrinsic step-like temperature dependence of the thermal conductivity of polymer colloidal crystals opens avenue path to fabricate colloidal superstructures with geometric constriction-controlled thermal transport properties. **Figure 4**a shows the fabrication of a multilayer colloidal crystals with each of three layers containing different PMMA particles (containing varying volume percentage of *n*BA) with a same diameter of ~ 420 nm but different transition temperatures (i.e., $T_{g,1} = 61$ °C, $T_{g,2} = 103$ °C, $T_{g,3} = 124$ °C).[46] Although the heat transfer will be enhanced for the individual layer after reaching its corresponding $T_g$, the thermal resistances or contacts between colloidal particles are unchanged in the unmolten layers and therefore the material retains a low effective thermal conductivity (Figure 4b). Each time the temperature exceeds the $T_g$ of a given layer, the effective thermal conductivity of the sample will experience a stepwise increase. With this approach, colloidal crystals with specified number of stepwise increases in its thermal conductivity can be designed, and the temperature range of the transition. By tuning the thickness of the layers, the degree of transition change in each step (or the increment of the thermal conductivity) can also be effectively controlled.

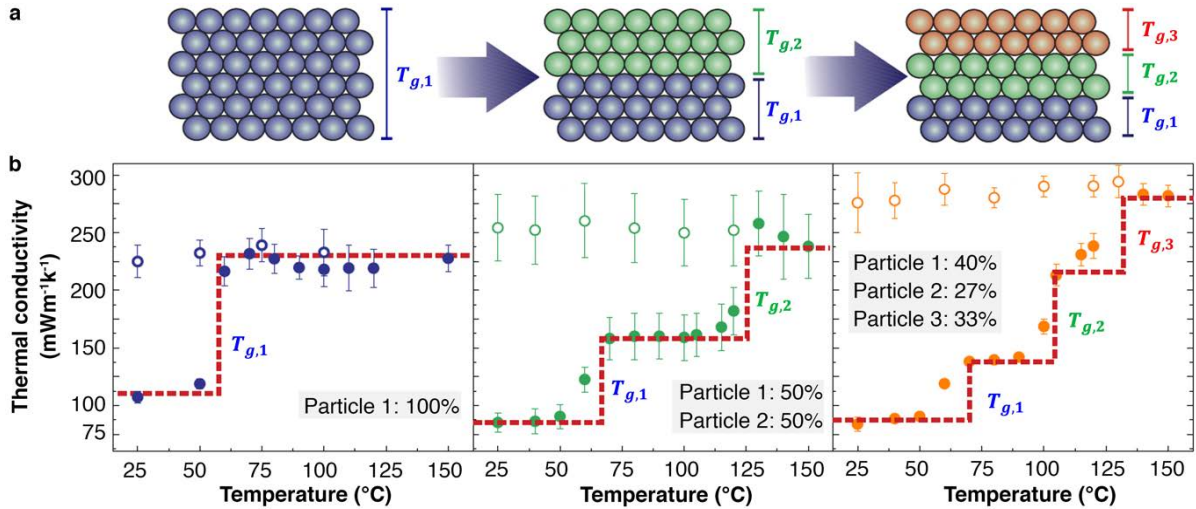

**Figure 4.** Thermal transport in colloidal crystals exhibiting multiple stepwise increase feature. (a) Schematic illustration of a colloidal monolith comprised of different PMMA particle layers and each layer has its own $T_g$. (b) The corresponding stepwise increase of the thermal conductivity due to the increasing temperature. Closed and open markers represent the heating and cooling cycle. Reproduced from Ref.[46]

***Particle size dependence***: Besides temperature, the colloidal assemblies can also be prepared from colloidal particles with different sizes to modify their thermal transport properties. **Figure 5**a shows how the thermal conductivity of a binary PS-based colloidal assemblies can be tailored by varying the percentage of the larger particles.[48] The two particles possess a diameter of $d_s = 243$ nm and $d_l = 306$ nm, respectively, with a size ratio $\eta_d$ of 0.79 (here, $\eta_d = d_s/d_l$). It is shown that the effective $\kappa$ of the colloidal assembly is very sensitive to the mixture ratio when the large particle volume ratio $\eta_v^l$ is below 20% or above 80%. Here, $\eta_v^l = V_l/(V_s + V_l)$, and $V_s$ and $V_l$ represent the volume of the small and large particles, respectively.

For intermediate mixing ratios $0.2 < \eta_v^l < 0.8$, the effective $\kappa$ drops to around 80% compared with that of the homo-particle colloidal assemblies.

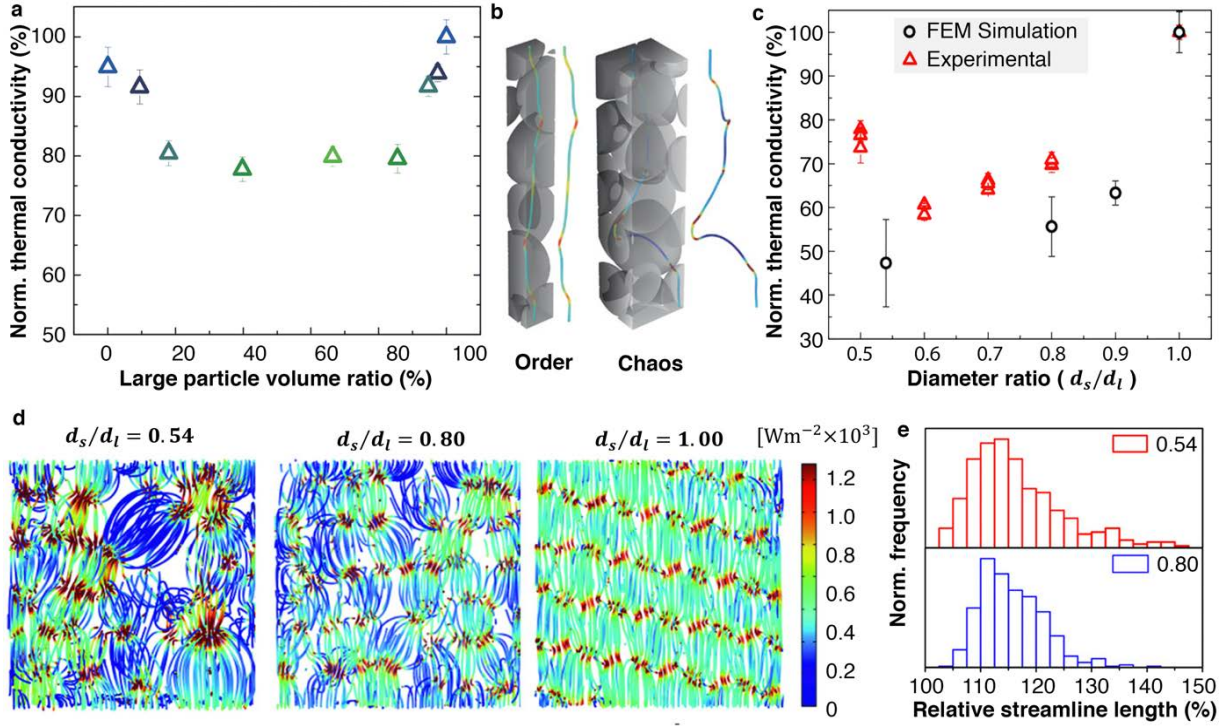

**Figure 5.** Thermal transport in colloidal assemblies with binary PS particle mixtures. (a) The normalized thermal conductivity (in relative to the sample with mono-large particles) as a function of the large particle volume ratio. (b) Schematic comparison of the heat flux streamline length in an ordered crystal and a disordered assembly. (c) The normalized thermal conductivity as a function of the size ratio at $\eta_v^l \sim 0.2$. (d) The heat flux densities of particle assemblies with varying size ratios. (e) Histograms of the streamline length at the size ratio of 0.54 (upper panel) and 0.80 (bottom panel). Reproduced with permission. [48] Copyright (2018), Wiley.

Two mechanisms are responsible for the suppressed thermal conductivity in a disordered colloidal assembly. One is the reduced density of the structure due to the non-close-packing, as the binary assembles can have as much as a 90% density reduction compared to the mono-particle crystals. The other is the increased thermal pathway or streamlines for the heat transport resulting from the size mismatch between adjacent particles. As illustrated in Figure 5b, the streamline is very straight and unperturbed in an ordered assemble, but is strongly bent and perturbed with the introduction of disorder (or different sized particles). According to the finite element method (FEM) modeling, the thermal transport pathways in the colloidal crystal are uniform and quite straight ($d_s/d_l = 1.0$ in Figure 5d), whereas the disordered colloidal assemblies possess highly distorted pathways. As illustrated in Figure 5e, the colloidal assembly with smaller size ratio contains higher number of long pathways, and their lengths can reach around 140% in relative to the length of the simulation box. Such observations suggest that the bigger difference between the particle diameters, the more likely it is that the heat flux streamlines will be distorted and elongated. As evidenced in Figure 5c, both experimental measurements and FEM simulations show a systematic increase of the thermal conductivity when the particle size mismatch diminishes, and a significant reduction of about 50% is observed for the scenario with a diameter ratio of about 0.54.

### *2.1.2 Hollow and Core-shell Particle-based Colloidal Crystals and Assemblies*

Hollow particles synthesized from silica or silicate are also being frequently utilized to fabricate colloidal crystals. Comparing with the solid polymer particle, the hollow particle colloidal crystals also provide effective opportunities to reach lower thermal conductivity. For instance, the hollow silica colloidal crystal shows a low thermal conductivity of 35 mWm$^{-1}$K$^{-1}$,[18] which is comparable with polymer foam-based thermal insulation materials.[36] Studies show that the particle geometry, the packing density and symmetry, and the interparticle bonding strength play an important role on the thermal transport of the hollow particle colloidal crystals.[36, 49]

Four different heat transport mechanisms have been proposed for the hollow particle colloidal crystals as illustrated in the **Figure 6**a,[41] including solid conduction, open- and closed-pore volume gaseous conduction, radiative transport, and convection. The solid conduction and the open- and closed-pore volume gaseous conduction are the major contributors. According to Figure 6b, the colloidal crystals with larger particles possess smaller thermal conductivity. For instance, the silica colloidal crystal with a particle size of 266 nm shows a thermal conductivity of about 80 mWm$^{-1}$K$^{-1}$ (in vacuum condition), which is more than two times higher than the counterpart with a particle size of 469 nm (~ 25 mWm$^{-1}$K$^{-1}$). The thermal conductivity also increases with the shell thickness, e.g., about 43% of increase (from 46 to 82 mWm$^{-1}$K$^{-1}$) in vacuum condition when the thickness increases from 14 to 40 nm (with a similar particle size of 266 nm). Such results suggest that the colloidal crystals can be built with a low thermal conductivity with larger particles but thinner shells at a price of stiffness reduction.[50] The influence on the heat transport from the open-pore volume is evidenced from the different thermal conductivity as measured in different environments. As shown in Figure 6b, the colloidal crystals exhibit a much lower thermal conductivity in vacuum condition than that under gaseous environment.

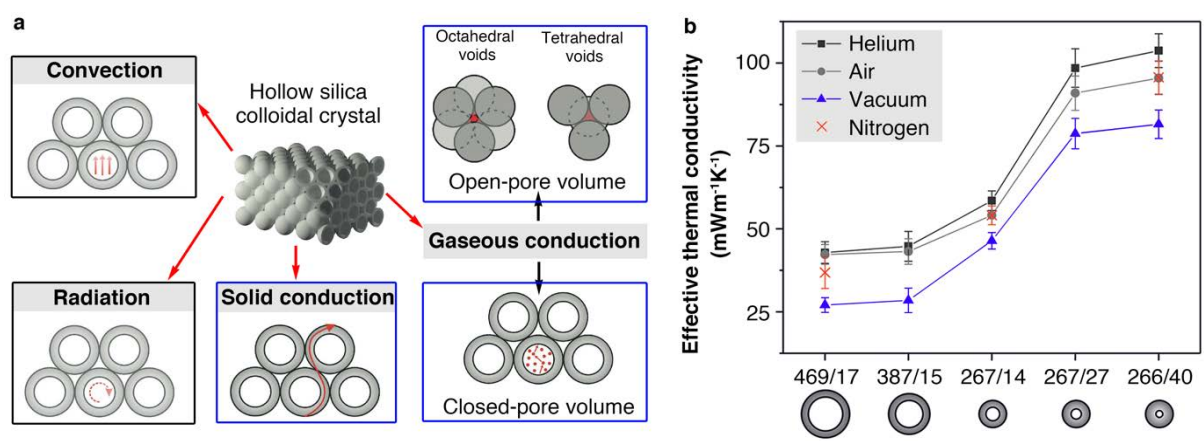

**Figure 6.** Thermal transport in colloidal crystal comprised of silica hollow sphere. (a) Schematic view of the thermal transport pathways. Reproduced with permission.[18] Copyright (2017), Wiley. (b) The thermal conductivity as a function of the diameter and shell thickness

of the hollow particle at 25 °C. The gaseous pressure is ~ 1000 mbar, and the vacuum is ~ 0.05 mbar. Reproduced with permission.[41] Copyright (2017), Wiley.

Considering the dominant solid conduction mechanism (Figure 6a), the thermal transport in the colloidal crystals can be further modified by changing the interparticle interactions, including altering the interaction strength and the number of contact points. Through calcination, strong covalent bonds can be introduced between adjacent hollow spheres, which reduces interfacial phonon scatting and leads to higher thermal conductivity. In the extreme scenario where there is only weak van der Waals (vdW) interactions, the colloidal crystal yields to the lowest thermal conductivity of only about 8 mWm$^{-1}$K$^{-1}$ in the vacuum condition (with a particle size and shell thickness of 469 nm and 17 nm, respectively).[41] Altering the interparticle interactions will also influence the temperature dependence of the thermal conductivity. For instance, a change in the temperature dependence from $\kappa \sim T^{0.5}$ to $\kappa \sim T^{0.4}$ is observed for the silica hollow sphere colloidal crystal after 500 °C and 950 °C calcination respectively.[49]

The hollow nature of the silica sphere also brings the opportunity to fabricate colloidal crystals with core-shell particles, which brings a phase change characteristic to the silica colloidal crystals. **Figure 7**a compares the thermal conductivity of colloidal crystals constructed with the diameter of PS-silica core-shell particles ranging from 269 nm to 479 nm (but an identical silica shell thickness of ~ 15 nm).[41] As is seen, the thermal conductivity increases significantly during the first heating cycle, which remains at the higher level during the cooling circle. Such results are analogous to that observed from the polymer particle colloidal crystals (Figure 3). Scanning Electron Microscope (SEM) images reveal that the PS leaks through the silica shells at higher temperature, which endows the thermal conductivity of the colloidal crystal a stepwise characteristic. As shown in Figure 7c, necks between adjacent particles are formed due to the PS leakage after the first heating cycle. Experimental results show that the PS leakage can be prevented by increasing the silica shell thickness. As compared in Figure 7b, the core-shell particle colloidal crystal with a shell thickness of 42 nm exhibits an almost linearly increasing thermal conductivity. According to the SEM images, there are no necks formed during the first heating circle.

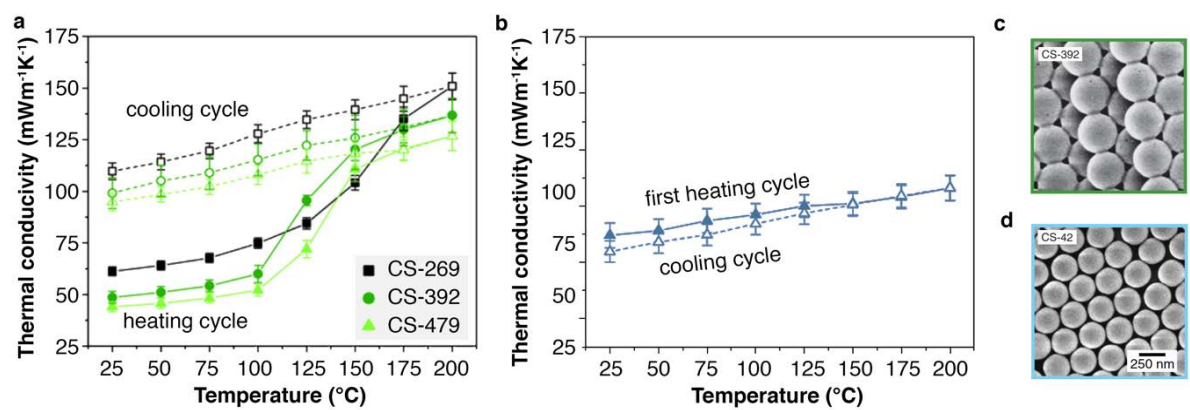

**Figure 7.** The influence of particle geometrical sizes on the thermal conductivity of the core-shell particles colloidal crystals. (a) The stepwise increase of the thermal conductivity when the temperature increases for the particle diameter of 269 nm, 392 nm, and 479 nm, respectively. All particles have an identical shell thickness of ~ 15 nm. (b) The temperature influence on the thermal conductivity of the core-shell particles colloidal crystals with a particle shell thickness of 42 nm. The diameter of the core-shell particle is 270 nm. A comparison of the SEM images of the core-shell particles after heating up to 200 °C for: (c) the particle with a diameter of 392 nm and shell thickness of 15 nm, and (d) the particle with a dimeter of 270 nm and shell thickness of 42 nm. Reproduced with permission. [41] Copyright (2017), Wiley.

In summary, investigations of the thermal transport properties of colloidal assemblies are still in their infancy. The colloidal particles can be fabricated using different materials, such as polymers, ceramics, metals, and semiconductors, and various shapes from solid sphere, hollow sphere, core-shell, to non-spherical shapes.[17] These are also known to possess inherent functional properties such as photonic and phononic properties, and with polymer materials, the phase transition of the constituent materials gives the colloidal assemblies unique temperature-dependent thermal conductivity. Specifically, different external stimuli may be used to trigger the transition, such as pH, solvents, light, electric or magnetic fields.[46] These varieties bring a huge possibility to construct multiphysical and multifunctional heat management materials based on colloidal assemblies (that can be integrated into thermal switches, transistors, or diodes), which is a fertile field for more research efforts in the near future.

### *2.1.3 Three-dimensional Foam*

Comparing with the colloidal crystals/assemblies, an analogous type of polymer-based 3D porous structure – polymer foam (also called nanocellular foam), has been extensively studied. The extremely low thermal conductivity (typically in the range of 30 – 40 mW/m-1K-1) of the polymer foams makes them ideal for thermal insulation applications.[51] The polymer nanocellular foams are normally prepared with nanoconfined voids (or channels with trapped gas) in a stable solid matrix, producing open-cell foams or closed-cell foams.[51a]

Similar as the colloidal crystals in Figure 6a, there are four heat transfer mechanisms involved in the nanocellular foams, including the contribution from radiation ($\kappa_r$), convection ($\kappa_c$), gas phase conduction ($\kappa_g$) and solid phase conduction ($\kappa_s$). According to the Grashof number ($Gr$), convection of air is enabled when ($Gr \geq 1000$), which corresponds to the void diameter larger than 10 mm. Thus, gas convection in the nanocellular foams makes no contribution to the heat transfer.[52] Meanwhile, the radiation contribution can be calculated from $\kappa_r = \frac{16}{3} \frac{n_i^2 \sigma T^3}{e(T)\rho}$. Here, $n_i$ and $\rho$ are the mean index of refraction and the apparent density of the specimen, respectively; $T$ is the temperature; and $\sigma$ is the Stefan-Boltzmann constant.

$e(T)$ denotes the specific Rossland mean extinction coefficient, which dependents on the constituent materials and affects the overall heat transfer of the nanocellular foams.[52] The solid phase conduction depends on the phonon transport, which can be described by the kinetic theory of gases, i.e., $\kappa_s = \frac{1}{3} C_p v_g \Lambda$, where $C_p$ and $v_g$ are the specific heat capacity and the mean group velocity of phonons; and $\Lambda$ is the phonon mean free path (MFP).[1a] The gas phase conduction originates from the collisions of gas molecules between themselves and with the solid walls, which is described by the kinetic theory model with a correction factor that accounts for the influence of viscosity.[53]

It is reported that the gas phase conduction contributes significantly to the effective heat transfer of the polymer foam (up to 70%),[54] which can be calculated from $\kappa_g = \kappa_0/(1 + K_n)$. Here $\kappa_0$ is the thermal conductivity of a bulk gas, and $K_n = \Lambda_g/d$ is the Knudsen number (with $\Lambda_g$ as the MFP of gas molecules and $d$ as characteristic dimension of the voids). It is clear from above relationship that the gas phase conduction decreases when the Knudsen number increases. In other words, the dimension of the voids can effectively change the thermal conductivity of the polymer foam. Using this concept, recent research has focused on the fabrication of polymer nanocellular foams with gas-filled pore (or void) sizes down to 100 nm, which can on the one hand prevent structural failure but on the other hand turn the gas into Knudsen flow.[55] In theory, the gas molecules will enter the ballistic regime and the gas phase conduction will be significantly suppressed when the pore sizes are on the order of the MFP of gas molecules (e.g., around 70 nm for air at ambient condition). Extensive results have shown an ultra-low thermal conductivity of the polymer nanocellular foam with pore sizes down to 100 nm. For instance, Rizvi et al successfully fabricated polypropylene foams using industrial-scale foam molding technique to achieve a low thermal conductivity of 29 mWm$^{-1}$K$^{-1}$ (for air) at a density of 0.074 gcm$^{-3}$.[56]

Overall, researchers have investigated the thermal transport of the polymer nanocellular foams with various constituent polymer matrices, different pore sizes, and different densities. Since there are already several recent reviews devoted to the thermal conductivity of nanocellular foams,[51] we will not focus on this group of structures in this review.

### *2.1.4 Nanolattice*

Besides the colloidal crystals/assemblies and 3D polymer foams, a recently reported 3D hierarchical structure – the nanolattice, also shows promising applications for thermal isolation. Nanolattices, also known as mechanical metamaterials, are prepared by high-precision additive manufacturing techniques,[57] whose certain mechanical properties are defined by their geometry rather than their composition.[58] Due to the size effect at nanoscale, the nanoscale mechanical metamaterials, opens up a new material-property design domain. Different techniques have been developed to fabricate nanolattices,[59] such as direct laser writing (DLW), self-assembly, genetic engineering. Nanolattices fabricated from different materials have been demonstrated, including carbon (**Figure 8**a),[23, 60] ceramic (Figure 8b),[61] metals or alloys,[62] and metallic glass.[63] Extensive research has already explored the mechanical performance of nanolattices, with excellent mechanical properties found, including high deformability, high recoverability, and ultrahigh stiffness.[64] For instance, the alumina nanolattices exhibit an excellent energy-absorbing capability, and the full recovery is observed even with over 50% compressive strain.[61a] The deformation mode of the hollow-tube $Cu_{60}Zr_{40}$ metallic glass nanolattices are found to be tailorable through the tube-wall thickness and temperature, i.e., shifting from a full shape recovery characteristic to ductile, and also brittle characteristics.[65]

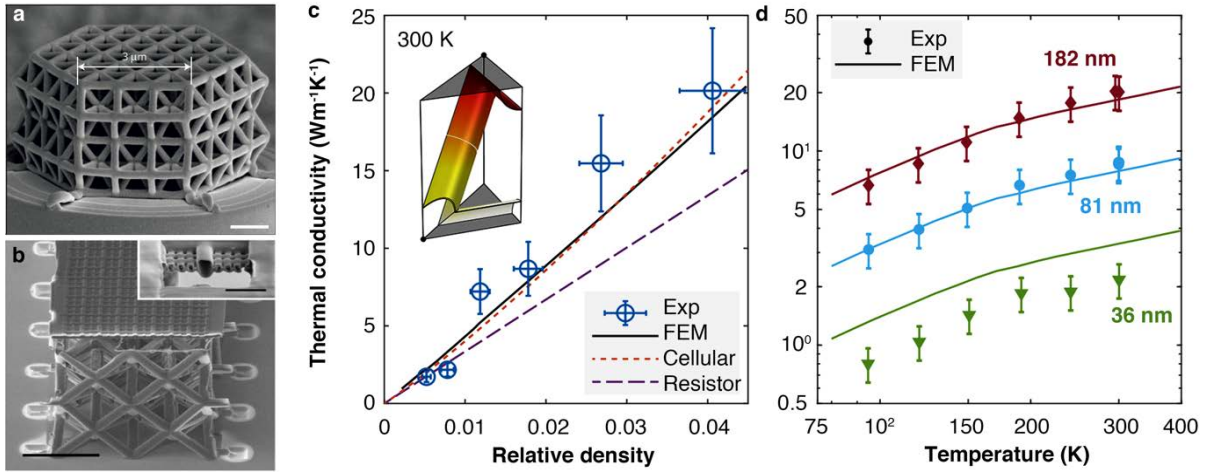

**Figure 8.** Thermal transport in nanolattice. (a) A carbon nanolattice fabricated by DLW and subsequent pyrolysis. Reproduced with permission. [23] Copyright (2016), Springer Nature. (b) The hollow-beam alumina lattice showing the octet-truss architecture. (c) The influence of density on the thermal conductivity of the alumina nanolattice. (d) The thermal conductivity of different alumina nanolattices at varying temperature. The nanolattice has a wall thickness of 36 nm, 81 nm, and 182 nm, respectively. (b)-(d) are reproduced with permission. [66] Copyright (2018), American Chemical Society.

Though significant understanding of the mechanical properties of nanolattices has been established, only one experimental work has probed the thermal transport properties of the alumina nanolattice. The relative density and the structural imperfections (such as thickness nonuniformities and beam waviness) are found to have a marked influence on the heat transfer in alumina nanolattices.[66] Figure 8c compares the experimentally measured thermal conductivity of the hollow-beam alumina nanolattices with the predictions based on the resistor[67] and cellular models[68] (both models are only dependent on the relative density of the structure). It is found that the measured thermal conductivity agrees well with the FEM simulation results, which are based on the classical effective medium theories, implying the dominant diffusive heat transport in the alumina nanolattice.

Figure 8d plots the temperature dependency of the thermal conductivity, from which we see that $\kappa$ increases as the temperature increases. This observation is consistent with that previously reported for amorphous oxide thin films.[69] The reasonable agreement between the FEM simulations and experimental measurements further suggest diffusion heat transport is dominant in the nanolattice. Overall, a low thermal conductivity, as low as 2 mWm$^{-1}$K$^{-1}$ is measured in the alumina nanolattice. It is expected that the thermal conductivity of the nanolattice can be further suppressed by creating multilayered walls (to promote interfacial phonon scattering). Given the rapid advances in additive manufacturing technology and the excellent mechanical performance of the nanolattices, it is of great interest to further exploit their thermal transport properties for thermal isolating applications.

## 2.2. Low Thermal Conductivity for Thermoelectric Device

Another application that demands low thermal conductivity is the thermoelectric (TE) device, which possesses great technological potentials in the energy sector. As mentioned previously, efficient thermoelectric devices require a high figure of merit ($ZT$), which is calculated from $ZT = S^2 \kappa_e \kappa_t^{-1} T$. Here, $S$ is the Seebeck coefficient; $\kappa_e$ and $\kappa_t$ are the electrical and thermal conductivity, respectively; and $T$ is the average temperature within the material. To achieve a high $ZT$, significant work has been conducted to prepare or design materials with suppressed thermal conductivity (but high electrical conductivity), such as core-shell NWs, phononic crystals, GeTe-based nanomaterials,[70] and SnSe-based nanomaterials.[71] There are extensive comprehensive reviews devoted to TE materials,[5] here we will only discuss the thermal conductivity of some representative 3D nanostructures, including nanowire arrays, inverse opals structures, and holey nanostructures or phononic crystals.

### 2.2.1 Nanowire Arrays

Si nanowire (NW) arrays are one of the nanostructures being frequently discussed for TE device with a low thermal conductivity (but a high electrical conductivity).[72] Besides Si NW arrays, metallic NW arrays have also been prepared, but targeting the usage as thermal interface materials with a high thermal conductivity. For their utilization in thermal interface materials, the NWs act as an expressway for the heat transport,[73] which are normally infiltrated with a polymer to form a composite structure. We will discuss the thermal conductivity of NW array-based composite in the following Sec. 4.4. NW arrays can be prepared via electrodeposition (**Figure 9**a)[21] or bio-template mask (Figure 9b).[74]

Bulk Si has been shown to have a poor ZT ~ 0.01 due to its high thermal conductivity (around 150 $Wm^{-1}K^{-1}$ at room temperature). Boukai $et$ $al$ reported that by varying the size and impurity doping levels of Si NWs, the ZT value of the Si NW can reach ~ 1 at 200 K, which is approximately 100-fold improvement.[75] These results demonstrate the promising applications of Si NWs in TE devices, and greatly promote the studies on the thermal transport properties of Si NW arrays.

Studies have shown that the thermal conductivity of Si NW arrays is highly dependent on their geometrical parameters (e.g., length and diameter), pore size, the polymer filler, and surface doping.[76] For instance, the discrete surface doping of Ge on Si NW arrays can induce 23% reduction in the thermal conductivity at room temperature compared to the undoped counterpart.[76d] As illustrated in Figure 9c, the thermal conductivity of the Si NW array decreases continuously with the increase of pore diameter and a low $\kappa$ around 1.68 $Wm^{-1}K^{-1}$ is observed at 300 K with a mean pore size around 12.35 nm.[77] The decreasing $\kappa$ results from enhanced phonon scattering at boundaries and nanopores.[78] Through the bio-template mask, recent studies show a much smaller $\kappa$ of the Si NW arrays around 0.5 ± 0.1 $Wm^{-1}K^{-1}$ along the axial direction over a temperature range of 300-350 K. These arrays have a high density and ordered structure (with a height of 30 nm).[74a] When the NW length (material thickness) increases to 100 nm, $\kappa$ increases to $1.8 \pm 0.3$ Wm$^{-1}$K$^{-1}$ (see Figure 9d). The increasing thermal conductivity with the length is commonly observed in 1D nanostructures[34a, 79] and 2D nanoribbons,[80] which can be explained from the perspective of the kinetic theory.[81] By embedding the same NW array into SiGe$_{0.3}$ matrix, a similar low thermal conductivity of $3.5 \pm 0.3$ Wm$^{-1}$K$^{-1}$ is measured in the composite structure in the temperature range of 300 – 350 K.[74b] Such low thermal conductivity of the Si NW array results from the enhanced phonon boundary scattering and the phonon confinement effect.[82]

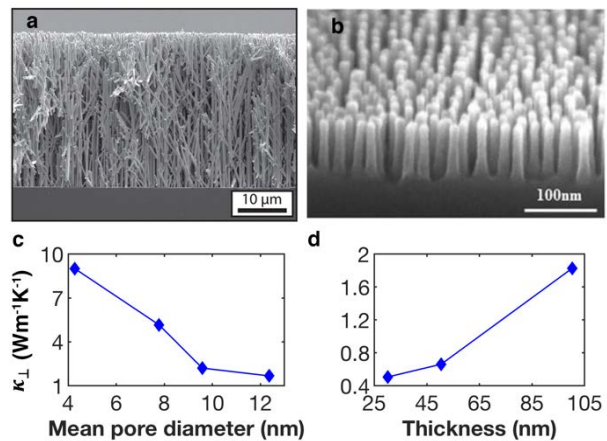

**Figure 9.** Thermal transport in NW arrays. (a) SEM images of the Cu NW array. Reproduced with permission.[21] Copyright (2015), American Chemical Society. (b) SEM image of Si NW array. Reproduced with permission.[74a] Copyright (2017), AIP Publishing. (c) The axial thermal conductivity of etched Si NW array as a function of the mean pore diameter at 300 K.[77] The diameter of the NW varies from 20 nm to 200 nm. (d) The axial thermal conductivity of the Si NW array as a function of the sample thickness in the temperature range of 300-350 K. The Si NW has a dimeter of 10 nm. Reproduced with permission. [74a] Copyright (2017), AIP Publishing.

Overall, the Si NW array has been shown with significantly suppressed thermal conductivity but unchanged electronic properties,[72b] and their thermal transport properties have been extensively studied, targeting TE applications.[83] Besides NW arrays, a lot of research has focused on the thermal conductivity of individual Si NW by investigating the impacts from a wide range of factors, such as surface roughness, geometrical size (diameter and length), temperature, defects (e.g., grain boundaries or stacking faults), and doping.[34a, 72b, 84] In addition to that, extensive studies have also been carried out on different Si crystal nanostructures, such as the porous Si nanomeshes (as discussed in the following Sec. 2.2.3), and the nanocrystalline bulk Si.[72c]

### 2.2.2 Inverse Opals Structures

The structural periodicity in the inverse opals (IOs) structures provides an effective avenue to tailor their heat transport properties for TE devices. Analogous to the colloidal crystals, the intrinsic optical properties of IOs enabled the fabrication of sensors based on color responses. Their highly porous nature with a large surface-to-volume ratio also make them ideal building blocks for the Li-ion batteries (e.g., the anodes or cathodes).[85] Generally, IOs are fabricated by first depositing the inverse opals materials into the interstitial spaces of a sacrificial opal template, and then removing/dissolving the opal template, which will thus form a 3D structure with interconnected spherical pores.

The heat transport path in the IOs can be mimicked by the interconnected hard spheres that occupy the interstitial sizes as schematically shown in **Figure 10**a,[19] and two controlling mechanisms are proposed for the thermal transport.[86] At the continuum length scale (the diffusive heat transfer), the porosity distorts the heat transfer pathway, which induces strong thermal resistance (similar to the example in Figure 5b for the colloidal crystals). At the micro/nanoscale, strong phonon scattering occurs at surfaces and grain boundaries, which remarkably suppresses the thermal conductivity of the material. Figure 10b compares the thermal conductivity of Si inverse opals with varying grain size ($s$) and shell thickness ($h$).

The Si inverse opals are fabricated from the 3D fcc silica opal template that is comprised of particle's diameter in the range of 300 to 640 nm.[86-87] As is shown, a low thermal conductivity (~ 1 Wm⁻¹K⁻¹) is observed for all examined Si inverse opals. Specifically, the temperature dependence of the thermal conductivity can be approximated as $\kappa \sim T^{1.8}$ in the low temperature regime, which deviates from the well-known $\sim T^3$ dependence of the heat capacity (for bulk). Such deviation is believed to result from coherent phonon scattering at grain boundaries that is frequency-dependent, similar to what is observed in nanocrystalline Si.[88]

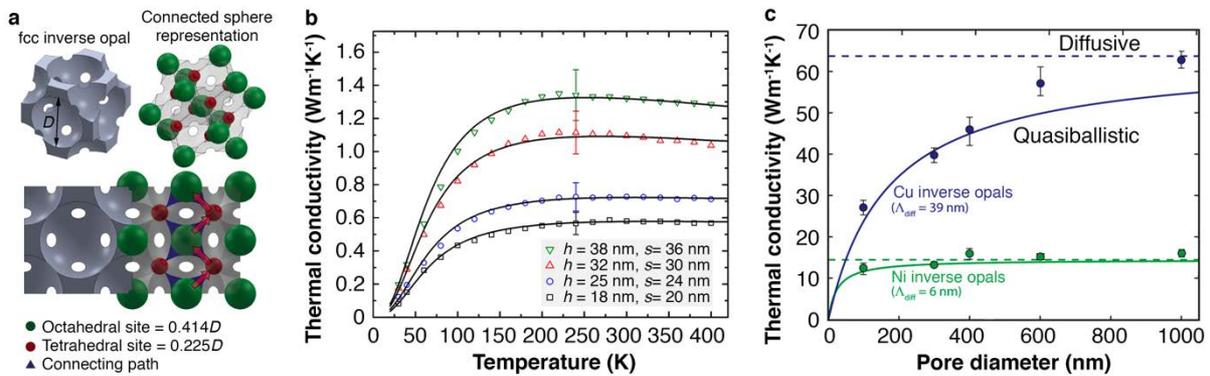

**Figure 10.** Thermal transport in inverse opals structures. (a) Schematic view of a fcc inverse opal unit cell (left). The octahedral and tetrahedral interstitial sites are highlighted by the space-filling spheres. The interconnected networks between the space-filling spheres (red) and the linkages (blue) represent the heat transfer pathway (red arrows). Reproduced with permission.[19] American Chemical Society. (b) The influence of temperature on the thermal conductivity of the inverse opals. The structures have different grain size (*s*) and silicon shell thickness (*h*). Reproduced with permission.[86] Copyright (2013), American Chemical Society. (c) The thermal conductivity of Cu and Ni close-packed inverse opals as a function of the pore diameter at room temperature. Reproduced with permission.[19] Copyright (2016), American Chemical Society.

The thermal conductivity of the inverse opals is highly related with the pore size, as expected from the primary mechanisms governing heat transfer discussed above. Figure 10c compares the thermal conductivity of Cu and Ni inverse opals measured at room temperature when the pore diameter changes from 100 nm to 1000 nm.[19] Basically, the thermal conductivity increases when the pore diameter increases, which saturates to the diffusive limit. In particular, the Cu inverse opals with the smallest pore diameter (~ 100 nm) is only about 43% of its counterpart with a dimeter of 1000 nm. According to the kinetic theory, the thermal conductivity is linearly related with the MFP of the energy carriers under the gray approximation, which is reasonable as the metals have a narrow electron MFP distribution with energy (e.g., 90% of electron thermal conductivity comes from electrons with MFPs between 22 nm and 41 nm for Au).[89] Since both surface scattering and internal scattering (e.g., electron-phonon, electron-grain boundary) affect the heat transfer in inverse opals, the effective electron MFP ($\Lambda_{eff}$) can be calculated based on Matthiessen's rule from $\Lambda_{eff} = \left(1/\Lambda_{diff} + 1/\Lambda_{surface}\right)^{-1}$. For the fcc inverse opals, the surface-limited MFP $\Lambda_{surface} \approx 0.233D$ (the ballistic limit), with $D$ as the diameter of the pore.[19, 90] Thus, the thermal conductivity of the inverse opals can be derived as a $\kappa_{IO} = \kappa_{IO,diff} \times 0.233D/(0.233D + \Lambda_{diff})$. According to the Drude-Sommerfeld model, the diffusive electron MFP is 39 nm for Cu and 6 nm for Ni at 293 K. Evidently, the thermal conductivity of the Cu and Ni inverse opals align with the diffusive limit as the effective MFP is close to the diffusive MFP for large pore diameter. As such, Ni inverse opals exhibit a diffusive-like thermal conduction for a diameter of around 100 nm as they have a short electron MFP.

Like the colloidal assemblies, the thermal transport properties of the inverse opals structures are still lacking of investigation. Given the large variety of materials (e.g., metals and semiconductors), there are great opportunities to design novel inverse opals-based heat management materials. Besides TE devices, inverse opals structures also exhibit promising applications for phase change materials used for thermal storage. For instance, by infiltrating Cu inverse opals with paraffin wax, the temperature rise in kWcm$^{-2}$ hotspots is suppressed by ~ 10% compared with the copper thin film heat spreaders.[91] In the thermofluidic system, metal inverse opals can be used as microfluidic heat exchanger with a high thermal conductivity. In general, inverse opals represent an important category of porous morphologies for enhanced interfacial transport in applications like electrochemical surfaces and microscale heat exchangers or suppressed heat transfer for TE devices, while their thermal conduction principles (e.g., the heat transfer across the interface) are still requiring extensive studies.

### 2.2.3 Holey Nanostructures

In recent years, the 3D nanostructure with regular vertically etched holes – holey Si structure or phononic crystals (as acoustic analogs of photonic crystals), have attracted extensive experimental and theoretical efforts. The periodic holes are usually introduced to the sample by etching.[92] In theory, the periodic holes will remarkably increase the surface-to-volume ratio and enhance surface phonon scattering, which will thus significantly suppress the thermal conductivity and benefit the applications in TE devices.[93] The majority of current studies have focused on the Si phononic crystals being prepared as 2D nanomeshes or films. A low thermal conductivity of around 1.9 Wm$^{-1}$K$^{-1}$ was reported for a holey silicon film, which is more than 20% reduction compared to the Si NW array device.[93] As illustrated in **Figure 11**a, a holey silicon nanostructure can be described by three geometrical parameters, i.e., the neck size ($n$), pitch length ($p$) and thickness ($L$). The hole diameter can be calculated from $d = p - n$.

The holey nature of the structure endows the phononic crystal with anisotropic thermal conductivity. Specifically, the cross-plane thermal conductivity can be tuned by the sample thickness and the in-plane thermal conductivity can be tailored by the hole diameter (or the neck size). As illustrated in Figure 11b, the cross-plane thermal conductivity exhibits an increasing tendency with the sample thickness. Such size-dependence can be described by the spectral scaling model that accounts for the phonon frequency dependence, i.e., $\kappa_z = \int_0^{\omega_D} \kappa_{z\infty}(\omega, n) \times \left(1 + \frac{\lambda_{z\infty}(\omega, n)}{L/2}\right)^{-1} d\omega$. Here, $\omega$ and $\omega_D$ are the phonon frequency and the Debye cut-off frequency, respectively; and $\kappa_{z\infty}(\omega, n)$ and $\lambda_{z\infty}(\omega, n)$ are the thermal conductivity and phonon MFP, respectively.[94] Similarly, the increase of the neck size results in a larger thermal conductivity (Figure 11c).[25, 92]

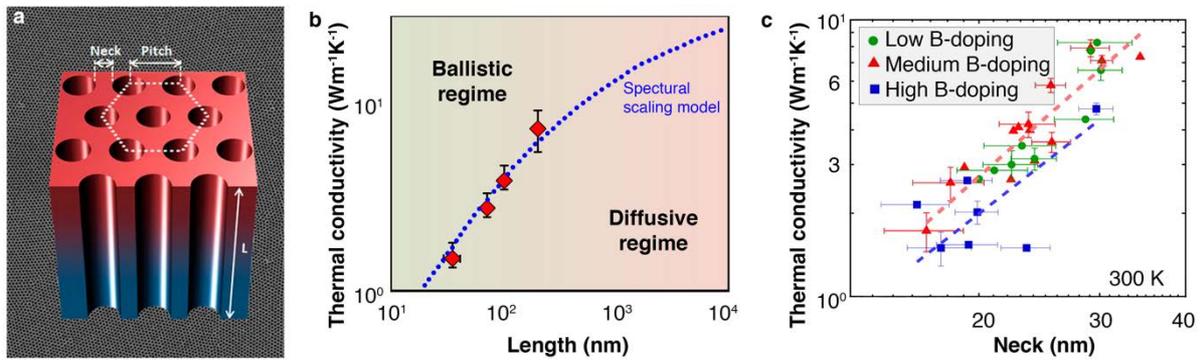

**Figure 11.** Thermal transport in holey silicon. (a) Schematic illustration of the holey silicon nanostructure. (b) The thermal conductivity of the holey silicon at different sample lengths. (a) and b are reproduced with permission.[94b] Copyright (2015), American Chemical Society. (c) The increasing thermal conductivity with the neck size. Reproduced with permission.[25] Copyright (2016), American Chemical Society.

Besides the dependence on the geometrical parameters, the thermal conductivity of the holey silicon exhibits a temperature dependence ($\kappa \sim T^2$) different from the classical temperature dependence ($\kappa \sim T^3$) in the temperature range of 20-60 K.[94b] Two mechanisms are proposed for such observation, one is the quantum confinement effect, which alters the phonon dispersion relation.[95] Another one is the enhanced role of the surface disorder and the boundary scattering, which is experimentally observed at room temperature.[96] The phonon confinement effect has been argued in literature, some studies show that the confinement effect is only significant for low temperature, e.g., around and below 6.4 K for Si NWs with a diameter of 10.8 nm.[97] While, other studies show that the phonon confinement occur even at room temperature.[98]

Overall, low thermal conductivity, sufficient electrical properties and good mechanical strength make the holey Si nanostructures or Si phononic crystals an ideal candidate for TE applications. Though it is widely accepted that the nanoscopic holes suppress the thermal conductivity, the underlying mechanism is still under argument. The phonon dispersion relations are potentially modified due to the presence of periodic holes in the nanostructure. From the perspective of coherent phonon transport, the thermal conductivity would be reduced due to the resulting phononic bandgaps and the reduced group velocities.[99] Several works have argued the role of coherent phonons on the thermal transport of holey nanostructures,[25, 96, 100] and the range of occurrence of the phonon interference is still an open question. In a recent work, Maire et al show the thermal conduction controlled by coherence in Si holey nanostructures over a relatively large temperature range, until the transition to purely diffusive heat conduction was observed at 10 K.[100a] With recent advances of nanofabrication technology, it is expected that the working temperature range of phononic crystals can be further broadened.

## 3. Three-dimensional Nanostructures for High Thermal Conductivity

At the opposite extreme of TE devices or thermal isolation, many electronic applications require materials to possess a high thermal conductivity for efficient heat removal during operation. In this regard, extensive works have been conducted to explore the nanomaterials with high thermal conductivity, especially for the thermal interface material (TIM) as used in modern electronic packing. Here we will discuss the recent progress on three categories of nanostructures pursuing high thermal conductivity, including the 3D architectures, metal-matrix composites with nanostructured fillers, and polymer composites with 3D nanostructured fillers. The thermal transport properties of polymer composites will be discussed separately in the following section due the large volume of relevant works in literature.

## 3.1. Hierarchical Structures for Thermal Management

### 3.1.1 Nanotube Arrays

Besides low thermal conductivity, the assembled or packed hierarchical nanostructures have also been prepared or proposed for high thermal conductivity applications, among which, the array structure is one of the simplest scenarios. Vertically aligned carbon nanotube (VACNT) arrays have received extensive attention. Due to the high thermal conductivity of individual CNTs, the VACNT arrays possess a high effective thermal conductivity parallel to the CNT axial direction, which make them ideal candidate as thermal interface material. Researchers have prepared TIMs based on either pure VACNT arrays or VACNT array-based composites (embedded in polymer). Almost all studies have shown that VACNT composites exhibit a decreased performance compared with their pristine array counterparts,[20] which is caused by the degraded CNT alignment, damping of phonon modes, and poor contact between CNTs and substrates. In this regard, a large volume of work has focused on the heat transport of pristine VACNT arrays. Typically, the total thermal resistance of a TIM can be determined from $R_{TIM} = R_{c1} + h_{blt}/\kappa_{int} + R_{c2}$, which correspond to three components (see **Figure 12**a), including the thermal contact resistances of the interfaces at the two ends ($R_{c1}$ and $R_{c2}$), and thermal resistance of the TIM ($h_{blt}/\kappa_{int}$).[20] Here, $h_{blt}$ and $\kappa_{int}$ are the thickness and the intrinsic thermal conductivity of the TIM, respectively. Therefore, to ensure effective heat removal, it is essential to minimize the thermal contact resistances at the two interfaces and enhance the intrinsic thermal conductivity of VACNT arrays.

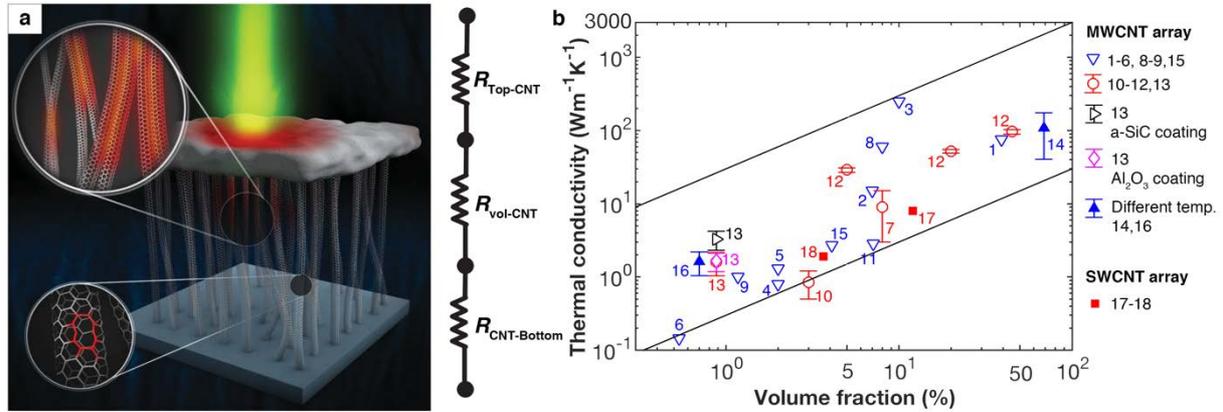

**Figure 12.** Thermal transport in vertically aligned CNT arrays. (a) Schematic illustration of a pure CNT array-based TIM. Reproduced with permission. [101] Copyright (2013), American Physical Society. (b) The influence of volume fraction on the intrinsic thermal conductivity of VACNT arrays, including multi-walled CNT (MWCNT) arrays (1 – Ref.[102], 2 – Ref.[103], 3 – Ref.[104], 4 – Ref.[105], 5 – Ref.[106], 6 – Ref.[107], 7 – Ref.[108], 8 – Ref.[109], 9 – Ref.[110], 10 – Ref.[111], 11 – Ref.[112], 12 – Ref.[113], 13 – Ref.[114], 14 – Ref.[115], 15 – Ref. [116], 16 – Ref. [117]), and single-walled CNT (SWCNT) arrays (17 – Ref.[118], 18 – Ref.[119]). Error-bars represent the range of the thermal conductivity measured under different conditions. Reproduced with permission.[101] Copyright (2013), American Physical Society.

The intrinsic thermal conductivity of VACNT arrays is highly dependent on the quality of the constituent CNTs and their packing density. Catalytic chemical vapor deposition (CVD) has been widely used to grow VACNT arrays from supported or floating catalysts, and many factors have been identified that affect their quality, such as the growth temperature, pressure, precursors, and the active catalyst.[120] The as-produced high-quality VACNT arrays exhibit a thermal conductivity in the order of 10 Wm⁻¹K⁻¹, which is better than most commercially available TIMs.[20] Various post-growth treatments have been proposed to enhance the heat transport in VACNT arrays, such as high-temperature annealing, microwave annealing, and coating. An earlier dedicated review comprehensively compared the intrinsic thermal conductivity of VACNT arrays reported before 2013.[101] Thus, this section will be limited to extend some of the results reported recently. Figure 12b summarizes the intrinsic thermal conductivity of VACNT arrays at different volume fractions. The dashed/solid lines represent the predicted intrinsic thermal conductivity of VCNAT array based on $\kappa_{int} = \phi \kappa_{cnt}$.[101] Here, $\phi$ is the volume fraction, and $\kappa_{cnt}$ is the thermal conductivity of an individual CNT. According to Figure 9b, some VCANT arrays exhibit an equivalent thermal conductivity of the individual CNT less than 30 Wm$^{-1}$K$^{-1}$, signifying other factors other than the CNT density degrade the thermal transport in the array structure. Overall, all reported thermal conductivities of the array fall well below the predictions based on the individual CNT with a thermal conductivity of 3000 Wm$^{-1}$K$^{-1}$.[121]

To date, considerable work has been carried out to investigate the thermal transport properties of VACNT arrays, which have focused on their intrinsic thermal conductivity. To facilitate their engineering applications, more studies are expected to investigate the total thermal resistance of the VACNT array-based TIMs, especially the interface heat transfer between VACNT array and the mating surface. Currently, it is still challenging to fabricate and transfer high quality VACNT arrays, which greatly affect the interfacial heat transfer.[20] It is shown that different agents (e.g., In, Cu, Au) can be used to bond CNT tips with mating surfaces to reduce the thermal contact resistance,[116, 122] however, it is unclear how reliable the bonding structure is under long-time thermal cycling.

### 3.1.2 Pillared Structures

Along with the CNT arrays, the high thermal conductivity of graphene has also enabled it as appealing building block for 3D hierarchical structures.[123] Pillared graphene is one of

WILEY-VCH

hierarchical structures that received extensive attentions recently, in which the graphene layer is supported by CNT pillars (**Figure 13**a and b). The CNT pillared graphene structures were initially proposed for the $H_2$ storage application,[22] and subsequently their electronic transport properties[124] and mechanical properties[125] have been studied. Although several papers have reported the successful synthesis of the pillared graphene,[126] their properties have rarely been studied experimentally, primarily due to the preparation and measurement difficulties. To date, only one study has investigated the thermal transport in CNT pillared graphene,[127] which however only measured the thermal resistance ($9 \times 10^{-10}$ $m^2KW^{-1}$) at the CNT-graphene junction and demonstrated the high heat dissipation capability of the structure. In comparison, several papers have investigated the thermal conductivity (either in-plane or cross-plane) of the pillared graphene through molecular dynamics (MD) simulations,[128] from which the major factors influencing their thermal transport properties have been identified, such as the junction configurations,[129] and the CNT length and density (or inter-CNT distance).[130]

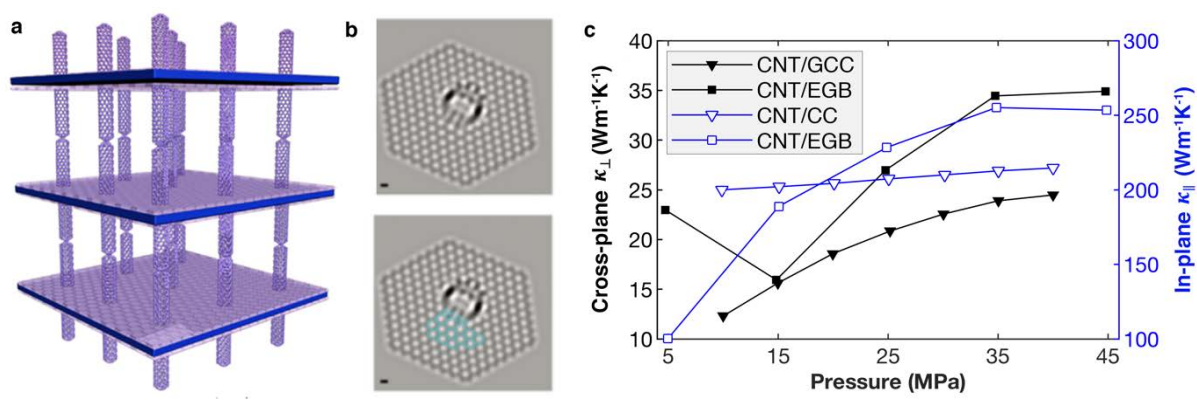

**Figure 13.** Thermal transport in pillared graphene. (a) Atomic configuration of a CNT pillared graphene. Reproduced with permission.[131] Copyright (2014), Elsevier. (b) Simulated STEM image of CNT and graphene junction areas. Reproduced with permission. [126b] Copyright (2012), Springer Nature. (c) The thermal conductivity of 3D carbon nanostructures

as a function of the hot-pressing pressure. CNT/EGB represents 3D CNT/exfoliated graphite block,[132] and CNT/GCC represents 3D CNT/graphite consolidated composite.[133]

Apart from the hybrid structures of the CNT-pillared graphene, there are also some other 3D carbon-based hierarchical structures being reported, as listed in **Table 1**, including the CNT/graphene oxide(GO) hierarchical structure, CNT/graphene hierarchical structure (annealed from CNT/GO at 1000 °C);[134] the 3D CNT network;[135] the CNT/ultrathin graphite foams;[136] and the 3D CNT intercalated graphite.[132-133] For the 3D CNT/GO hierarchical structured film with 15wt% of CNTs, an ultrahigh in-plane thermal conductivity of 1388.7 Wm$^{-1}$K$^{-1}$ was reported at 1000 °C, which decreases at either higher or lower weight fraction of CNTs.[134] In comparison, the 3D CNT network exhibits a cross-plane $\kappa$ of about 19.8 Wm$^{-1}$K$^{-1}$, which decreases with increasing transverse interconnected CNTs.[135]

**Table 1.** The thermal transport properties of 3D CNT based hierarchical structures.

| 3D hierarchical structures | Preparation method | $\kappa$ (Wm$^{-1}$K$^{-1}$) | $R$ (cm$^2$KW$^{-1}$) | Year |
|---|---|---|---|---|
| CNT/graphite foam (1.8wt% graphite and 0.8wt% CNT) | CVD and hot-pressing | 4.1 ± 0.3 | - | 2015[136] |
| CNT/exfoliated graphite block (EGB) | CVD and hot-pressing | 38 ($\kappa_\perp$); 290 ($\kappa_\parallel$) | - | 2016[132] |
| CNT/graphite consolidated composite (GCC) | CVD and hot-pressing | 24.3 ($\kappa_\perp$); 211.5 ($\kappa_\parallel$) | - | 2014[133] |
| 3D CNT/GO (15wt% CNTs, at 25 °C) | Vacuum filtration | ~ 3 ($\kappa_\parallel$) | - | 2017[134] |
| 3D CNT/graphene (15wt% CNTs, at 1000 °C) | Vacuum filtration and annealing | 1388.7 ($\kappa_\parallel$) | - | 2017[134] |

| 3D CNT network | CVD | | ~19.8 ($\kappa_\perp$) | 0.67~0.78 | 2018[135] |
|---|---|---|---|---|---|
| CNT pillared graphene | CVD | | - | $9 \times 10^{-6}$ | 2018[127] |

For the 3D hierarchical structures prepared through a hot-processing technique, their thermal conductivities are highly dependent on the growth time, growth density, hot-pressing temperature, and pressure.[132-133] A higher hot-pressing temperature leads to a higher cross-plane and in-plane thermal conductivity.[132] Figure 13c shows the cross-plane ($\kappa_\perp$) and in-plane ($\kappa_\parallel$) thermal conductivity under different hot-pressing pressure for the 3D CNT/GCC and the 3D CNT/EGB. Due to the high in-plane thermal conductivity of the graphite plate, the in-plane thermal conductivity of the CNT/GCC or CNT/EGB is nearly one order of magnitude greater than the cross-plane thermal conductivity. In general, the in-plane and cross-plane thermal conductivities increase as the hot-pressing pressure increases, which results from a denser structure and fewer voids at higher pressure. The decreasing of $\kappa_\perp$ for CNT/EGB (from 5 MPa to 15 MPa) is likely to be a result of the interplay between interlayer thermal contact resistances and in-plane thermal conduction in the graphite plate.[132] In particular, the graphite plates are expected as not aligned completely perpendicular in the pressing direction under low pressure, and thus their high in-plane thermal conductivity contributes to a higher overall thermal conductivity. At higher pressure, the graphite plates are better aligned, which counterbalances the reduced thermal resistance between graphite layers and leads to a temporary reduction to $\kappa_\perp$.

### 3.1.3 Honeycomb

It is worthy to mention that researchers have also proposed to construct 3D nanostructures based on 2D nanomaterials such as graphene[137] and BN,[137a] which form a regular honeycomb structure. The initial honeycomb structure was proposed by Karfunkel and Dressel in 1992.[138] Thereafter, there has been a sustained research on honeycomb structures, covering the structural stability and varieties,[139] mechanical properties,[140] electronic properties,[141] thermal transport properties,[139a, 141-142] and gas storage applications.[24] However, majority of these works are based on theoretical calculations or MD simulations, and the successful synthesis of the honeycomb structure was not reported until 2016 (**Figure 14**a).[24] Besides the carbon 3D honeycomb structures, recent studies also discussed the boron nitride-based honeycomb.[143] To the best of our knowledge, the thermal conductivity of honeycomb structure has not been experimentally measured. According to the theoretical calculations or MD simulations, the carbon honeycomb exhibits anisotropic thermal transport properties, i.e., a larger axial thermal conductivity than the transverse thermal conductivity. These in silico studies have also shown marked influences on the thermal conductivity of the honeycomb structures from various factors, such as the cell structures (or morphologies),[141, 142c] mechanical strain,[142b] and temperature.[142a] Figure 14b illustrates the specific axial and lateral thermal conductivity of a carbon honeycomb structure with varying density as obtained from density functional theory (DFT) calculations. Comparing with the thermal conductivity of graphene ($\sim 4000$ $Wm^{-1}K^{-1}$), the honeycomb structure has a considerably smaller $\kappa$ by a factor of 15 to 45.[139a]

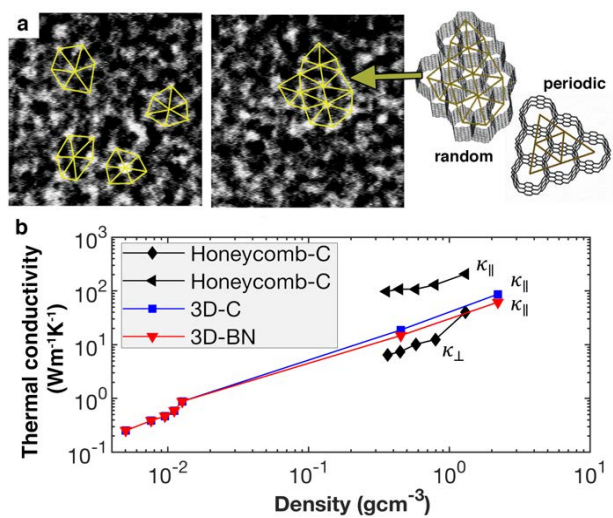

**Figure 14.** Thermal transport in carbon honeycomb structure. (a) Transmission electron microscopy (TEM) images of a carbon film showing the honeycomb structure. Reproduced with permission. [24] Copyright (2016), American Chemical Society. (b) The DFT calculated thermal conductivity of a carbon honeycomb structure,[139a] and the experimental measurements for 3D graphene and 3D BN foam (denoted as 3D-C and 3D-BN, respectively).[137a]

A few works have also studied the thermal conductivity of 3D foam structures constructed from 2D nanostructures. One recent work shows that the 3D-C and 3D-BN (as prepared via CVD method) exhibit an increasing cross-plane thermal conductivity ($\kappa_\perp$) when the density of the structure increases (Figure 14b). Comparing with the $\kappa_\perp$ of graphene and hBN paper (between 1-5 $Wm^{-1}K^{-1}$),[144] the 3D-C and 3D-BN exhibit a significantly high $\kappa_\perp$ of $86 \pm 10$ $Wm^{-1}K^{-1}$ and $62 \pm 10$ $Wm^{-1}K^{-1}$, respectively. Such remarkable enhancement is resulted from the covalent connections at the interfaces, which enables efficient phonon transport.[137a]

Overall, there are extensive works being conducted on CNT arrays, however, current studies on the thermal transport properties of other complex 3D hierarchical nanostructures (such as pillared graphene, honeycomb) are still very limited. For the CNT/graphite composites, the introduction of CNTs are found to enhance the cross-plane thermal conductivity, whereas they will impair the heat transport in the in-plane direction. In terms of the pillared structures and honeycomb stuctures, their thermal conductivity is still yet to be measured experimentally. There are several difficulties around these hierarchical structures, for instance, it is very challenging to fabricate and characterize 3D hierarchical nanostructures. The experimental measurement is also facing extensive challenges due to the structural complexities.

## 3.2. Metal-matrix Composites with Nano-structured Fillers

Metal-matrix composites (MMCs), comprising a contiguous metal matrix and reinforcements (filler materials), are widely used as large-scale structural materials.[145] Within the MMC, the metal matrix acts as a compliant support for the reinforcement fillers, and commonly used metals include Al, Ag, Cu, Mg, and Ti. The reinforcement fillers are used to tailor the properties of interest. Conventional MMCs are reinforced by ceramic fibers and particles in order to achieve a high-temperature resistance, high mechanical performance (e.g., specific strength and specific elastic modulus), and a low coefficient of thermal expansion (CTE). Besides structural materials, MMCs also have broad applications in the thermal management field for miniaturized electronic devices. Compared to their monolithic metal counterparts, MMCs are expected to possess a high thermal conductivity and a small CTE. Benefiting from their ultra-high thermal conductivity and low CTE, carbonaceous nanomaterials (e.g., CNT or graphene) have been widely utilized as a filler for MMCs. To illustrate, at room temperature, the multi-walled CNTs (MWCNTs) exhibit a thermal conductivity over 3000 $Wm^{-1}K^{-1}$ and a CTE of -2.5 $ppmK^{-1}$.[1a, 146] Whereas, pure Cu has a $\kappa$ of around 400 $Wm^{-1}K^{-1}$ and a CTE of 17 $ppmK^{-1}$.[147] Several papers have comprehensively reviewed the fabrication/processing approaches, mechanical properties, tribological properties, corrosion properties, electrical properties, and applications of various MMCs.[145, 147-148] This section will outline the thermal conductivity of various MMCs as reinforced by CNTs.

### 3.2.1. Copper-matrix

Cu-matrix composites are the most extensively studied MMCs due to their wide applications in electrical and electronic applications. However, current studies show negligible enhancement or even degradation of the thermal conductivity of the Cu-MMCs when CNT fillers are used, compared to the thermal conductivity of pure Cu.[149] Several factors have been proposed to be responsible for such observation, including the intrinsic compatibility (or wettability) between CNTs and Cu, interfacial bonding, and dispersion.[150] These factors are closely related to the fabrication or processing techniques. Most of the current studies have discussed the volume (or weight) fraction dependence of the thermal conductivity, which however show divergent results.

**Figure 15**a compares some results that report a decreasing thermal conductivity when the volume fraction of CNT increases,[151] and some others that reported an initial increasing tendency of κ within a small volume fraction.[149, 152] Several factors may be responsible for such divergent observations, such as processing techniques; the dispersion/agglomeration and alignment of CNTs (CNT has a very low thermal conductivity in the transverse direction, ~5.5 Wm-1K-1);[153] interfacial bonding at CNT/Cu interface; and the geometrical parameters (length and diameter) and defects of CNT.[151c] The CNT/matrix interfaces act as a thermal barrier that causes strong lattice phonon scattering, resulting in degradation of the overall κ.[149] It is observed that there are extensive defects (including defects in CNT and grain boundaries in Cu matrix) and voids formed in the Cu-MMCs, and the high aspect ratio of CNT also leads to severe entanglement and kinks (Figure 15b-c).[151a] These facts promote phonon scattering and lead to a reduction in κ. Several studies have reported an improved thermal conductivity with surface treatments (e.g., with Mo-coating[154] or chemical treatment[152]), whereas the enhancement is very limited. It is shown from theoretical calculations (based on the mixture rule) that the super-aligned CNTs can aid the heat transfer in the Cu-MMC, however it is still lacking experimental validation.[155]

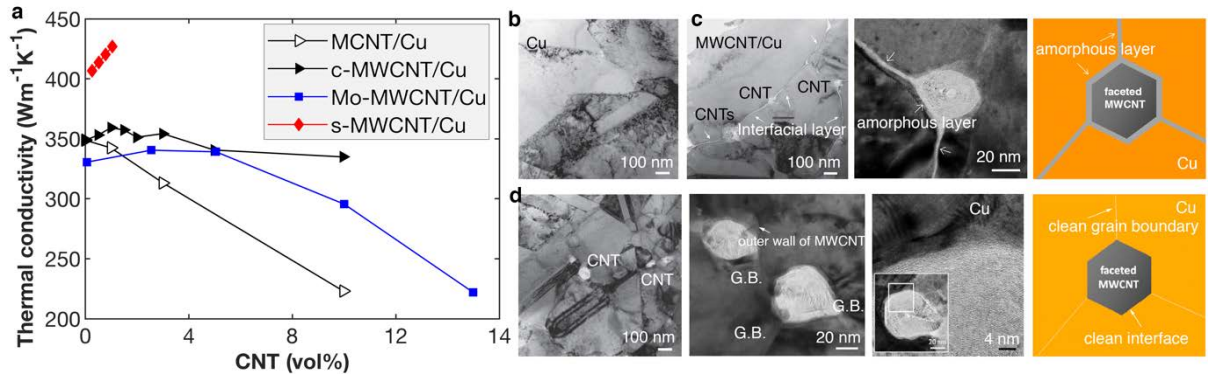

**Figure 15.** Thermal transport in Cu-MMCs reinforced by CNTs at room temperature. (a) The thermal conductivity of different Cu-MMCs at varying volume fraction of CNT, including the Cu matrix with pristine MWCNT and chemically treated MWCNT (denoted as MWCNT/Cu and c-MWCNT/Cu, respectively),[152] with Mo-coated MWCNT (M-MWCNT/Cu),[154] and with super-aligned MWCNT (s-MWCNT/Cu, theoretical calculation).[155] (b) TEM images of sintered Cu. (c) TEM images of pristine MWCNT/Cu (left), the interface between MWCNT and Cu (middle), and the schematic view of the interface. (d) TEM images of chemically treated MWCNT/Cu (left), the interface between MWCNT/Cu (middle), and the schematic view of the interface. (b)-(d) are reproduced with permission. [152] Copyright (2012), Elsevier.

Very few works have explored how the thermal conductivity of Cu-MMCs will change under different temperatures. One early work shows that the thermal conductivity fluctuates around 250 Wm-1K-1 for a CNT volume fraction of 5 vol.% in the temperature range of 298 K to 550 K,[149] indicating a relatively temperature independent behavior. However, a recent work shows that the thermal conductivity decreases from 310.5 Wm-1K-1 to 296.7 Wm-1K-1 in the temperature range of 20 °C to 500 °C.[27] Overall, to achieve a high thermal conductivity, the Cu-MMCs are supposed to contain well-aligned and well dispersed CNT fillers. The interfacial bonding of CNT/matrix and the structural damage of CNTs are also crucial influential factors, which are mainly unknown in the current studies.

### 3.2.2. Other Metal-matrices

In addition to Cu, CNTs have also been introduced to other metal matrices such as Ag, Al, Ni, Ti, Mg and alloys, but the composite thermal conductivities have not been fully investigated. Similar to Cu-MMCs, both enhancement and degradation effects are observed in the heat transfer due to the introduction of CNTs. For example, κ of the Ag-MMC with chemically treated MWCNTs deceases significantly from 430 Wm-1K-1 to ~ 175 Wm-1K-1 when the volume fraction of CNT increases from zero to 6%,[156] which is most likely as a result of the covalent bonds between CNT and the matrix that impair the electron conduction. By comparison, the Ag-MMC with non-covalent functionalized MWCNT (that improves the wettability of CNT with the matrix) exhibits an increasing κ with the CNT content (**Figure 16**a). Some early work shows that the CNT/Al MMC exhibits an increasing κ at low volume fraction which then decreases at higher volume fraction.[157] Interestingly, the experimental results show that the alloy-matrix MMCs normally show an increasing thermal conductivity with increasing CNT volume fraction, such as the Cu-Cr composite,[158] Cu-Ti composite,[159] and 70Sn-30Bi alloy composite (with Ni-coated CNTs).[160] Such enhancement is explained from the perspective of the wettability of CNT, i.e., a thin interface layer that reduces the Kapitza resistance formed in the alloy matrix.

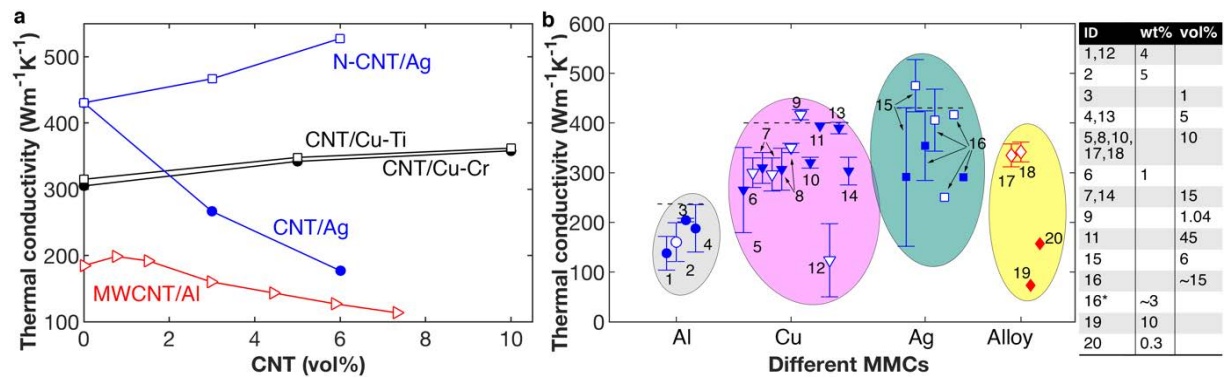

**Figure 16.** Thermal conductivity of various MCCs reinforced by CNTs at room temperature. (a) The thermal conductivity exhibiting a divergent relationship with the volume fraction of CNT for CNT/Cu-Cr,[158] CNT/Cu-Ti,[159] CNT/Ag (with covalent functionalization),[156] N-

CNT/Ag (with non-covalent functionalization),[156] MWCNT/Al (produced by spark plasma sintering, the weight fraction is converted to volume fraction based on a density of 1.8 gcm3 for MWCNTs and 2.7 gcm3 for Al).[157] (b) The thermal conductivity of different MMCs reinforced by CNTs, including Al-MMCs (1 - Ref.[161], 2 - Ref.[157], 3 - Ref.[162], 4 – Ref.[163]), Cu-MMCs (5 – Ref.[149], 6 – Ref.[151a], 7- Ref.[154], 8 – Ref.[152], 9 – Ref.[155], 10 – Ref.[164], 11 – Ref.[150], 12 – Ref.[165], 13 – Ref.[27], 14 – Ref. [166]), Ag-MCCs (15 – Ref.[156], 16 – Ref.[167]), and alloy-MCCs (17 - Ref.[158], 18 – Ref.[159], 19 – Ref.[160], 20 – Ref.[147]). Dashed lines denote the thermal conductivity of monocrystalline metals. The error bars represent the variation of the thermal conductivity in the studied weight/volume fraction range. The solid markers represent a decreasing thermal conductivity within the examined weight/volume fraction, and the open markers represent an initial increasing and then decreasing thermal conductivity.

Figure 16b summarizes the reported thermal conductivity of MMCs reinforced by CNTs with their monolithic metal counterparts. Since the measurement methods vary in literature, and the MMCs contain different CNT volume or weight fractions, their thermal conductivities can only be qualitative compared. Compared to their monolithic metal counterparts, almost all types of CNT reinforced MMCs show a smaller thermal conductivity. In the MMC, the heat transfer relies on both electrons and phonons (i.e., $\kappa = \kappa_e + \kappa_p$), and the electron heat transfer plays the dominant role. Experimental results show that the addition of CNTs significantly disturb and block the electron transportation path, and this results in a decreasing κ when the filler loading increases.[157] In all, the CNT fillers are found to reduce the CTE for the MMCs, while high thermal conductivity has not yet been achieved. The investigation of the thermal transport properties of MMCs is still at a very early stage. The thermal conductivity of the MMC depends on many factors, such as the intrinsic heat transfer in the metal matrix and the nanofiller, and the interfacial thermal resistance. However, it is challenging to characterize the interfacial bonding between the metal matrix and the nanofiller, and the structural damage of the nanofiller during sample preparation is hard to be identified. It is also difficult to control the dispersion and alignment of nanofillers during sample preparation. Extensive experimental, theoretical and computational effort is still required to understand how the nanofiller can be used to enhance the heat transfer in MMCs.

## 4. Polymer Composites with Nanostructured Fillers for High Thermal Conductivity

As the common thermal interface materials (TIMs) in the electronic packing industry, polymer composites have received extensive interests from both scientific and engineering communities.[168] They also have broad applications as underfill materials, organic substrate materials, and in flexible electronic devices. To ensure efficient and effective thermal management, polymer composites are expected to possess a high thermal conductivity. As such, a large amount of research has focused on preparation of polymer composites with a high thermal conductivity by adding different types of highly (thermally) conductive fillers. Theoretically, there are many factors that constrain the thermal conductivity of the polymer composite, including the intrinsic thermal conductivity of the filler, the filler/filler and filler/matrix interface resistances, the dispersion and alignment of the filler, and the geometrical parameters of the filler (e.g., size, shape, and surface area).[169] Owing to the absence of heterogeneous interface, 3D nanostructured filler is the ideal option to reduce the filler/filler interfacial thermal resistance with a relative low filler loading. This section is intended to outline the recent progress in using different types of 3D nanostructured fillers in polymer composites.

## 4.1. Carbon Nanotube-based 3D Nanostructured Fillers

CNT has been reported with a thermal conductivity of ~3,000-3,500 $Wm^{-1}K^{-1}$ at room temperature along the axial direction.[121, 170] Extensive efforts have been devoted to incorporate CNT into polymers to improve their thermal conductivity. Researchers have reported a thermal conductivity of CNT/polymer composite reaching 10 $Wm^{-1}K^{-1}$ at a CNT loading less than 5wt%.[171] The common problem with CNT fillers is that they will improve the electric conductivity of the composites, which can be a problem when electrical insulation is required. To resolve this issue, different surface modifications have been proposed, such as fluorination, which could on the one hand achieve similar enhancement to $\kappa$ compared with the pristine CNT while retaining a good electrical insulation characteristic of the composite.[172] Several earlier papers already contain a wealth of information on the thermal transport properties of CNT reinforced polymer composites, covering the influential factors like the CNT geometries, alignment, dispersion, weight/volume fraction, and surface modifications.[173] As such, this section will only limited to the recent studies on the thermal transport properties of polymer composites with 3D nanostructures/networks based on CNT.

The implementation of 3D network filler can be dated back to 2005 when Zhang et al.[174] demonstrated a 3D network structure from the in situ growth of CNTs from the interlayer of clay. Thereafter, such method has been widely accepted and applied to construct 3D hybrid network structures using two or multiple fillers, such as graphene (or graphite) nanoplatelets/CNTs,[175] graphene oxide/CNTs,[176] boron nitride (BN)/CNTs,[177] SiC/CNTs,[178] and BN/AlN.[179] Based on the nature of the bond between CNT and other fillers, the following discussions will first discuss the thermal transport properties of polymer composites with weakly bonded 3D nanostructures (a common synergistic network), followed by the polymer composites with 3D nanostructured (a hetero-structure) fillers with chemically bonded interfaces.

### *4.1.1 Synergistic Network*

A synergistic effect in a polymer composite with two or more fillers arises when the increase in thermal conductivity of the composite is greater than the sum of the increase in thermal conductivity for each individual component. Benefiting from their ultra-high aspect ratio (length over diameter) and thermal conductivity, CNTs have been popularly adopted to form synergistic networks with other fillers, in which the synergistic networks are normally generated through melting and mixing binary or more type of fillers directly with the polymer. The synergistic effect ($f$) can be quantified as $f = \frac{\kappa_{AB} - \kappa_p}{(\kappa_A - \kappa_p) + (\kappa_B - \kappa_p)}$,[180] where $\kappa_A$ and $\kappa_B$ are the thermal conductivity of the polymer composite with filler A and B, respectively; $\kappa_p$ and $\kappa_{AB}$ are the thermal conductivity of the pristine polymer and the composite with the AB synergistic network, respectively. The effective synergistic network is formed when $f$ is larger than 1, and greater $f$ suggests stronger synergistic effect. Two-dimensional nanomaterials, such as BN nanosheet (BNNS) and graphene (nanosheet or nanoplatelets), have been commonly used to form the synergistic networks with CNT.[177, 180-181] For BNNS, its electrical insulation characteristic can interrupt the electrically conductive network of CNTs in the polymer composites, and retain the electrical insulation characteristic of the polymer.[182]

To optimally use the high intrinsic thermal conductivity of the filler in the polymer matrix, a percolation network of the filler is necessary, which relies heavily on the dispersion, aspect ratio, and content (volume fraction) of the filler.[183] Various additional factors are responsible for the synergistic effect of the 3D structured networks, such as the alignment and content of each filler, and the percolation threshold of individual fillers.[177, 181, 184] **Figure 17**a plots the $\kappa$ as a function of the filler loading for high-density polyethylene (HDPE) composites containing different types of fillers.[184] For the HDPE with mono-fillers, i.e., only expanded graphite (EG) filler, or only CNT filler, $\kappa$ exhibits a nearly linear increase tendency with the increase of filler loading. Such relationship is generally described by the Nan model, i.e., $\frac{\kappa_e}{\kappa_m} = 1 + \frac{\Phi_f AR}{3} \frac{\kappa_c/\kappa_m}{AR+P}$,[185] which is derived from the Maxwell-Eucken model[186] with the consideration of interfacial thermal resistance. Here, $P = \frac{2R_\kappa \kappa_m}{d} \frac{\kappa_c}{\kappa_m}$; $\kappa_e$ and $\kappa_m$ are the effective thermal conductivity of the polymer composite and the intrinsic thermal conductivity of polymer matrix, respectively; $AR$ and $\Phi_f$ are the aspect ratio and volume fraction of the filler, respectively; $R_\kappa$ is the interface thermal resistance; $\kappa_c$ is the thermal conductivity of the filler; and $d$ is the diameter of the filler. From Figure 17a, the $\kappa$ of the HDPE composite with 10wt% of EG increases linearly with the CNT content, similar as that observed from the EG/HDPE with increasing EG content. Such results indicate that the synergistic effect is not activated. In comparison, the addition of $\sim$ 0.5wt% of CNTs for the 15wt%EG/HDPE and 20wt%EG/HDPE composites triggers a sudden increase to the effective thermal conductivity, signifying the synergistic effect.

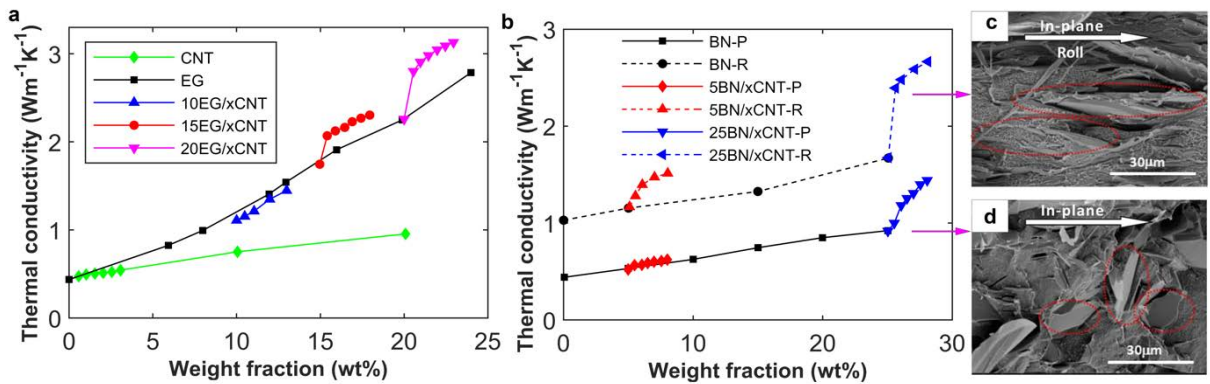

**Figure 17.** Thermal transport in HDPE composites with CNT-based synergistic networks. (a) The thermal conductivity of HDEP composites with single CNT fillers, single EG fillers, and a binary CNT and EG fillers. Reproduced with permission.[184] Copyright (2017), Elsevier. (b) The thermal conductivity of HDPE composites prepared by different techniques. R and P represent the hot rolling and hot-pressing, respectively. The samples contain single BN fillers, and a binary BN and CNT fillers. The dispersion states of the HDPE composites (25wt%BN

and 3wt%CNT) prepared by: (c) the rolling technique, and (d) the hot-pressing technique. (b)-(d) are reproduced with permission.[177] Copyright (2018), Elsevier.

The fabrication technique is found to have a profound effect on the synergistic network. Figure 17b illustrates the content dependence of the $\kappa$ for HDPE composites molded by hot-pressing (P) and hot rolling (R) (including BN and CNT fillers).[177] As is seen, the samples molded by hot rolling exhibit a much higher thermal conductivity, and the synergistic effect is activated at a much lower content of BN (~ 5 wt%).[177] According to the SEM images (Figure 17c and 17d), the hot rolling is able to arrange BN and CNTs along the rolling direction, which results in better filler alignment and more thermally conductive paths.

### 4.1.2 Segregated Double Network

Since the synergistic networks are normally generated through melting and mixing/blending binary or more types of fillers directly with the polymer, the enhancement to the heat transfer is still limited. To obtain interconnected networks with a lower filler loading, researchers proposed the fabrication of segregated double networks in the composites,[30, 187] i.e., the polymer particles are embedded with one conductive filler before being coated with another conductive filler. **Figure 18**a compares the thermal conductivity of PS composites that are prepared through three different procedures with a binary filler of CNT and 3.5vol% graphene nanoplatelets (GNPs), including a randomly dispersed network, the segregated network, and the segregated double network (Figure 18b).[30] As is seen, the thermal conductivity of the PS composites behaves differently with the volume fraction. For the sample with randomly dispersed network, $\kappa$ shows a gradual increase with no sharp change as the MWCNT content increase and no synergistic effect is observed. For the sample with segregated network, a decreasing $\kappa$ is observed, which is resulted from the interrupted contacts of GNPs/GNPs due to the introduction of CNTs. In comparison, the sample with segregated double network shows a significant increase in $\kappa$ with a small content of CNT (~ 1.5vol%). Such observation suggests that the synergistic double network can significantly enhance the heat transfer in the polymer composite compared with other 3D networks.

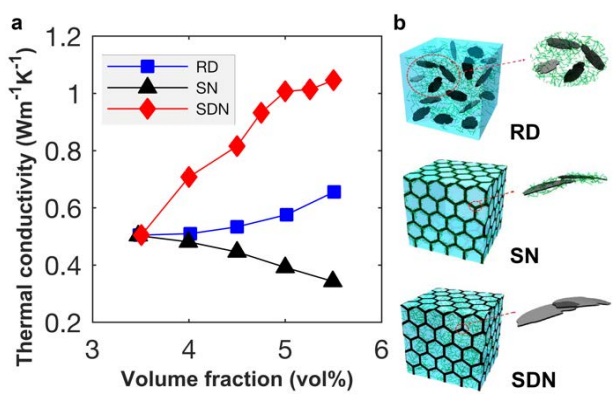

**Figure 18.** Thermal transport in PS composites with synergistic networks. (a) The distinguish changing tendency of the thermal conductivity of the PS composites (containing 3.5wt%GNPs) as prepared with a randomly dispersed network (RD), a segregated network (SN), and a segregated double network (SDN).[30] (b) The corresponding schematic illustration of the different polystyrene composites. (a) and (b) are reproduced with permission.[30] Copyright (2017), American Chemical Society.

### 4.1.3 Hetero-structure

In contrast to the weak interactions between fillers in the synergistic network, extensive efforts have also been attributed to establish 3D hetero-structures based on CNTs, within which the filler/filler interfaces are chemically bonded. The covalent bonds that bridge the multi-filler interfaces are expected to further reduce phonon scattering and achieve a more efficient heat transfer in the polymer composites. The hetero-structures are normally prepared by growing or grafting CNTs on the surfaces of other structures (e.g., carbon fibers[188]) or use CNT as skeletons to grow other low-dimensional nanostructures (e.g., $MoS_2$, graphene,[189] and BN nanosheet[190]). Different from the synergistic network, the hetero-structure allows a good control of the alignment of fillers, which endows the polymer composites with a controllable anisotropic thermal conductivity, i.e., a high cross-plane thermal conductivity along the CNT axis, and a low in-plane thermal conductivity in the transverse direction.

**Figure 19**a depicts the variation of the thermal conductivity of epoxy (EP) composites with different filler contents, including single CNT filler, hetero-structured $MoS_2$/CNT filler, and hetero-structured $MoS_2$/graphene/CNT filler. The hetero-structure is prepared by growing $MoS_2$ and graphene on surface of CNTs via a hydrothermal method.[189] As is seen, for the filler loading less than around 20wt%, $\kappa$ increases almost linearly with the filler content. According to the Foygel model,[191] the effective thermal conductivity of the composites can be estimated from $\kappa_e = \kappa_0 (\Phi_f - \Phi_c)^{t(a)}$, where $\kappa_0$ is the effective thermal conductivity of the filler network; $t(a)$ depends on the aspect ratio of the filler; and $\Phi_c$ is the volume fraction of the percolation threshold. Note that the Foygel model assumes percolating networks in the matrix with random distributed fillers. With the fitted values from the experimental measurements, the thermal contact resistance $R_c$ is then calculated from $R_c = \left( \kappa_0 L \Phi_c^{t(a)} \right)^{-1}$.[192] It is estimated that the $R_c$ for CNT/EP and $MoS_2$/CNT/EP are around 2.17-$3.98 \times 10^7$ KW$^{-1}$ and $1.9 \times 10^7$ KW$^{-1}$, respectively, which are much higher than the $MoS_2$/graphene/CNT/EP ($8.3 \times 10^6$ KW$^{-1}$). This enables higher heat transfer efficiency in the MoS2/graphene/CNT/EP composite than its counterparts.

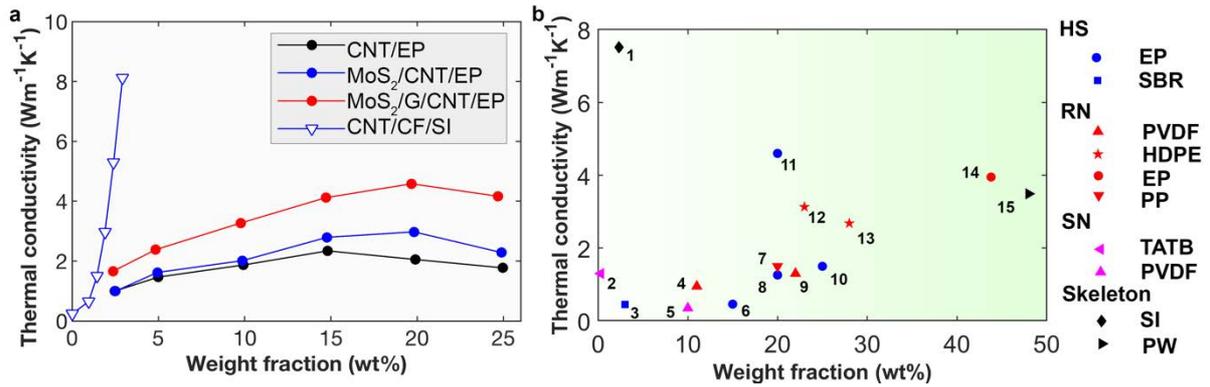

**Figure 19.** Thermal transport in the polymer composites with hetero-structured fillers. (a) Cross-plane thermal conductivity of EP composites with different fillers (including CNT filler, MoS₂/CNT filler, and MoS₂/graphene/CNT filler), and the silastic (SI) composites with carbon fibers (CFs) with 35 μm long aligned CNTs grown on their surface. (b) The highest thermal conductivity and corresponding weight fraction using CNT hetero structure fillers or networks (1-Ref.[188]; 2- Ref.[187b]; 3- Ref.[193]; 4- Ref.[176]; 5- Ref.[194]; 6- Ref.[195]; 7- Ref.[181]; 8- Ref.[196]; 9- Ref.[182b] ; 10- Ref.[190]; 11- Ref.[189]; 12- Ref.[184];13- Ref.[177]1;14- Ref.[182a] ;15- Ref.[197]). The marker shapes present the polymer base type, and the colors present the filler structures: the blue is the hetero-structured fillers (HS); the red is the random synergistic network (RN); the magenta is segregate network (SN); and the black is the skeleton network (Skeleton).

Other larger scale skeletons have also been used to grow CNTs on their surfaces to prepare hetero-structured fillers. For instance, the vertically aligned carbon fibers (CFs) with CNTs grown on their surfaces can increase the $\kappa$ of silastic (SI) composite to 8.14 $Wm^{-1}K^{-1}$ with a total carbon load of only 2.86wt%.[188] A recent paper reported the growth of CNT on a Cu foam, which successfully bridges the holes in the foam, and enables a higher $\kappa$ of 3.49 $Wm^{-1}K^{-1}$ in the paraffin composite.[197] Comparing with the synergistic networks, the preparation of hetero-structures is more complicated for industrial scale-up.

Figure 19b summarizes the recent works on the polymer composites with CNT-based 3D networks. For comparison, the polymer composites with randomly dispersed pre-modified CNT fillers are also listed, including the polymer functionalization,[193, 198] AlN doping,[199] and metal nanoparticles (NPs) deposition.[195] Surface functionalization are introduced in the purpose of reducing the thermal resistance at the CNT/polymer matrix interface. From Figure 19b, although the percolation networks are presented in the composites, majority of the prepared composites still exhibit a $\kappa$ lower than 2 Wm$^{-1}$K$^{-1}$, and usually a large filler loading (over 20wt%) is required to obtain a higher thermal conductivity.

It is noted that nearly all studies have emphasized the relationship between the thermal conductivity and the filler content. Unfortunately, it is challenging to establish a uniform theoretical model that can be used to describe and predict the thermal conductivity of the polymer composites. The difficulties arise from the complexities of the heat transfer in the polymer composites, which is related with the filler network, filler structure or morphology, filler dispersion, intrinsic thermal conductivity of the filler, and the interfacial thermal resistance (or phonon scattering) at the filler/filler interface and filler/matrix interface. Although different structures/networks have been explored, there is still a significant gap between the effective thermal conductivity of the polymer composites and the intrinsic thermal conductivity of nanostructured filler. Thus, a fundamental challenge that deserves further investigation is how to minimize the interfacial thermal resistance. Different functionalization strategies have been proposed for the nanofiller, which however normally degrades the intrinsic thermal conductivity of the filler.

## 4.2. *h*BN-based 3D Nanostructured Fillers

The high thermal conductivity, electrical insulation and low coefficient of thermal expansion, make inorganic ceramics another popular type of filler for insulator polymer nanocomposites.

Hexagonal boron nitride (hBN) is one of the most extensively investigated ceramic fillers for polymer composites. There are four common structures of hBN, including 0D NPs, 1D nanotubes (NTs), and 2D nanosheets (NSs), and 3D nanostructures (e.g., foam and network). Xu et al.[169] briefly reviewed the thermal conductivities of polymer composites reinforced by these BN nanostructures. A very recent review also concisely discussed the heat transport in the polymer composites with 2D BNNSs.[200] As such, this section will be limited to the thermal conductivity of polymer composites with 3D BN nanostructures or networks.

### 4.2.1 hBN Nanotubes

Similar to CNT, hBN nanotubes (BNNTs) possess high thermal conductivity ($\sim 350$ Wm$^{-1}$K$^{-1}$)[201] and high aspect ratio, which make them ideal for building percolation networks in polymer composites. Extensive research has investigated the thermal conductivity of BNNTs reinforced polymer composites, with an emphasis on the influences from the volume (or weight) fraction and temperature. The big gap between the axial and lateral thermal conductivity of the BNNT lead to the polymer composites (with aligned BNNTs) exhibiting anisotropic thermal transport properties, i.e., a higher thermal conductivity along the tube axis than that across the tube axis.

The temperature dependency of the thermal conductivity varies with the polymer matrix, which is related with the glass transition temperature ($T_g$) of the polymer matrix. For disordered nanostructures, the thermal conductivity increases with temperature when $T < T_g$, which is caused by the enhanced phonon transmission at the interfaces.[202] For the crystalline phase, $\kappa$ decreases with temperature due to the Umklapp phonon scattering. As illustrated in **Figure 20**a, the in-plane thermal conductivity of the cellulose nanofiber (CNF) composites increases with temperature before reaching $\sim 60$ °C,[201, 203] suggesting that the thermal transport of the amorphous polymer dominates the heat transport. In comparison, only a slight increase is observed for the epoxy composite with hBNNTs.[204]

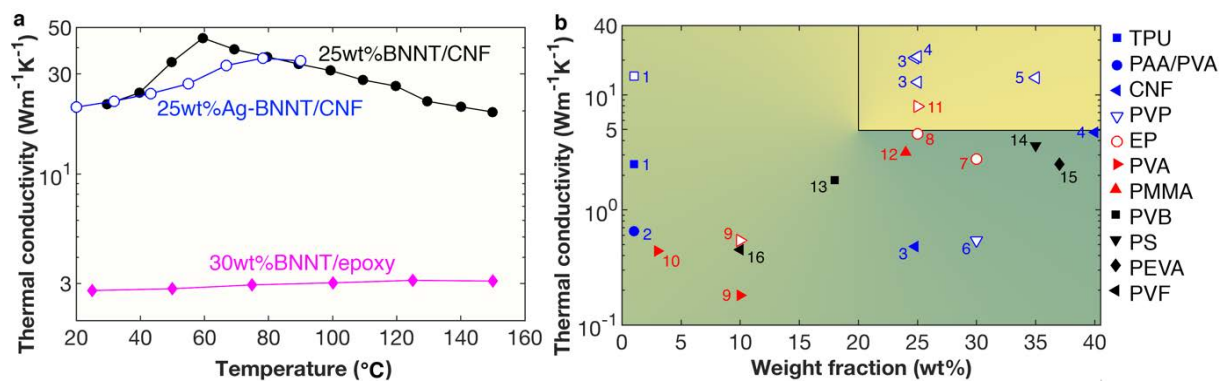

**Figure 20.** Thermal conductivity of polymer composite reinforced by hBNNTs. (a) The temperature dependency of the cross-plane thermal conductivity ($\kappa_\perp$) for the 25wt% BNNT/CNF,[201] 25wt%Ag-BNNT/CNF,[203] and 30wt%BNNT/epoxy composites.[204] (b) The thermal conductivity of different polymer composites reinforced by BNNTs, including thermoplastic polyurethane (TPU, 1 − Ref. [205]), polyacrylic acid (PAA)/polyvinyl alcohol (PVA) polymers (PAA/PVA, 2 − Ref. [206]), CNF (3 − Ref. [203], 4 − Ref. [201], 5 − Ref. [172]), polyvinylpyrrolidone (PVP, 6 − Ref. [207]), EP (7 - Ref. [204], 8 − Ref. [201] ), PVA (9 - Ref. [208], 10 − Ref. [209], 11 - [201]), polymethyl methacrylate (PMMA, 12 - Ref. [210]), polyvinyl butyral (PVB, 13 - Ref. [210] ), polystyrene (PS, 14 - Ref. [210]), polyethylene vinyl alcohol (PEVA, 15 - Ref. [210]), and polyvinyl formal (PVF, 16 - Ref. [209]). The open and solid markers represent the in-plane and out-of-plane thermal conductivity, respectively.

The weight (or volume) fraction exerts a marked influence on the thermal conductivity of the polymer composite. Both in-plane and cross-plane thermal conductivity are found to increase with the increase of weight fraction.[204-205] However, for the CNF composite, the in-plane thermal conductivity decreases with the weight fraction when the BNNTs content is less than 25wt% due to the lack of effective connections (or percolation networks) at low content of CNF.[201, 203] Different models have been developed to interpret the relationship between the thermal conductivity of the composite and the filler loading, e.g., the effective medium theory model that incorporates the interface resistance,[211] the Lewis-Nielsen model,[212] and the nonlinear model developed by Foygel *et al*.[191] Assuming the percolation critical power law, the best fitted cross-plane thermal conductivity of the CNT composite is around 60.2 $Wm^{-1}K^{-1}$, which is comparable to the range of 55-170 $Wm^{-1}K^{-1}$ for vertically aligned BNNT films. Such observation suggests the crucial role of the thermal resistance at the BNNT/matrix interface.

Figure 20b compares the thermal conductivity of various polymer composites reinforced by BNNTs. It is seen that the thermal conductivity of the polymer composite exhibits a large variation even the filler loading is the same. Such discrepancy primarily originates from the different alignment of the BNNTs and also the interfacial resistance of BNNT/polymer as determined by the preparation approach. For instance, using the hot-pressing technique, the BNNT/PMMA achieves a $\kappa$ of 3.16 $Wm^{-1}K^{-1}$ at 24 wt% of BNNTs,[210] which is 5-fold higher than the BNNT/PVP(polyvinylpyrrolidone) fibers fabricated from electrospun (about 0.54 $Wm^{-1}K^{-}$ at 30 wt% of BNNTs).[207] According to Fu et al.,[203] a low weight fraction of 0.199% of Ag NPs reduces the interfacial resistance of 25wt% BNNT/CNF from 1.7 $\times 10^{-9}$ $m^2KW^{-1}$ to 0.73 $\times 10^{-9}$ $m^2KW^{-1}$, which increases the $\kappa$ from ~ 12.9 $Wm^{-1}K^{-1}$ to 20.89 $Wm^{-1}K^{-1}$. Overall, high thermal conductivity (> 5 $Wm^{-1}K^{-1}$) of polymer composites with BNNTs have been reported, which however normally requires a high filler loading (over 20wt%).

### 4.2.2 Three-dimensional BN Networks

As aforementioned, 3D interconnected network is an effective way to enhance the heat transfer in polymer composites while keeping a relative low filler loading, which is benefited from the minimized interfacial phonon scattering between fillers. Different approaches/methods have been developed to construct 3D nanostructures based on BNNS or BNNT.[213] For instance, Zeng et al developed an ice-templated fabrication method,[32] which produces an ordered 3D honeycomb-like network (**Figure 21**a). Wang and Wu proposed to fabricate 3D BNNS network via multiple layer-by-layer assembly using 3D melamine foam (MF) (Figure 21b).[28] Through a two-step CVD method, Xue et al. synthesized a BN cellular architecture made of interconnective nanotubular hBN (Figure 21c).[29]

The temperature dependence of $\kappa$ for polymer composites varies with both the polymer matrix and the 3D BN networks. Take the epoxy composite with nacre-mimetic 3D BN network as an example, its thermal conductivity in the parallel direction to the lamellar layer (cross-plane) appears sensitive to the temperature, which increases from 6.07 $Wm^{-1}K^{-1}$ to 8.41 $Wm^{-1}K^{-1}$ when the temperature increases from 20 °C to 100 °C. However, in the perpendicular direction (in-plane), $\kappa$ is insensitive to temperature.[214] Comparing with the composite with randomly dispersed BNNS, a stronger temperature dependency of $\kappa$ is observed for the composite with 3D BN networks. Though some works report a slight reduction of the thermal conductivity at higher temperature, such as the polydimethylsiloxane (PDMS) composite with aligned and interconnected BNNS.[215] A general increasing thermal conductivity with increasing temperature is observed for the 3D BN networks reinforced polymer composites, which is resulted from the enhanced interfacial phonon transmission at the filler/matrix interface. Similar as the composites with BNNTs (Figure 20a), $\kappa$ exhibits a declination after the temperature is over a certain threshold value, e.g., around 70 °C for epoxy composite.[31, 216] Such observation is explained from the perspective of the aggravated phonon Umklapp scattering of the crystalline BN filler, which outweighs the decreased Kapitza resistance, and thus impairs the thermal transport in the composite.[1a]

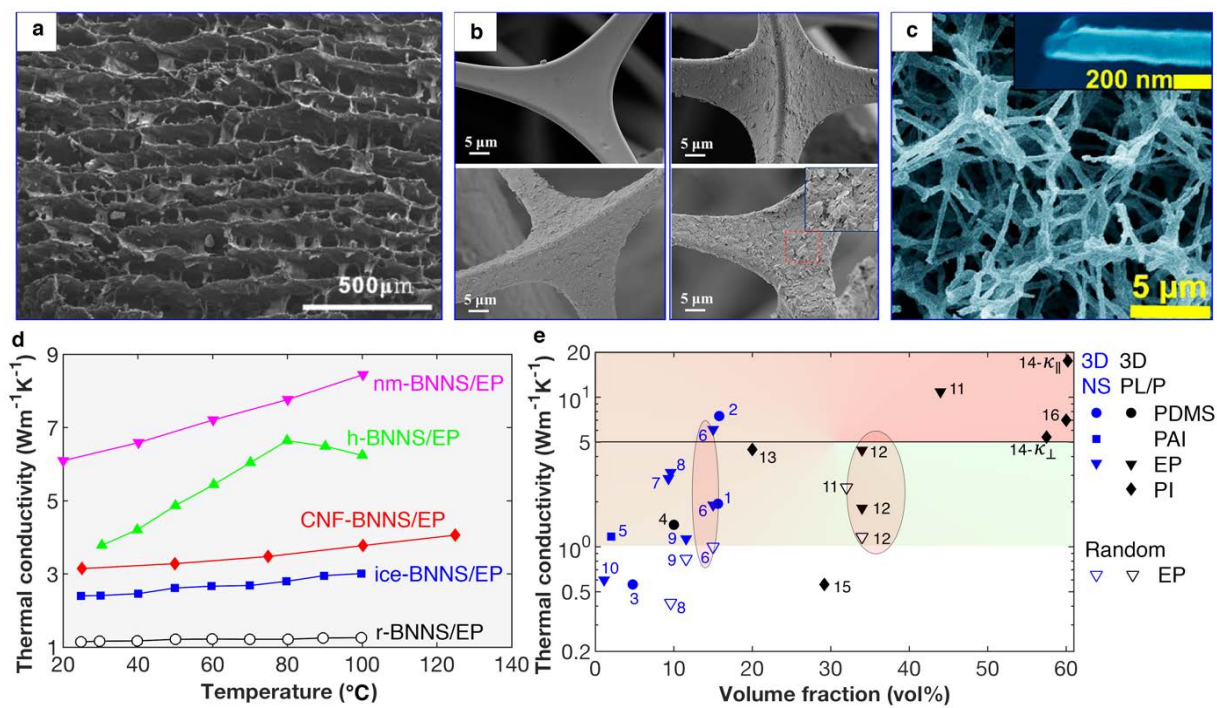

**Figure 21.** Thermal transport in polymer composites with 3D hBN networks. (a) SEM images showing the 3D-BNNS aerogels prepared from ice template. Reproduced with permission. [32] Copyright (2015), Wiley. (b) SEM images of MF foam with BNNSs assemblies. Reproduced with permission. [28] Copyright (2018), Elsevier. (c) SEM images of 3D BN cellular architecture made of interconnective nano-tubular hBN. Reproduced with permission. [29] Copyright (2017), American Chemical Society. (d) The thermal conductivity of epoxy composite reinforced by 3D BN networks, including the randomly dispersed BNNS (r-BNNS/EP, 9.29 vol% BNNS),[32] ice-templated assembly (ice-BNNS/EP, 9.29 vol% BNNS),[32] CNF supported (CNF-BNNS/EP, 9.6 vol% BNNS),[217] nacre-mimetic (nm-BNNS/EP, 15 vol% BNNS),[214] and hierarchically ordered (h-BNNS/EP, 31.8 vol% BNNS).[31] (e) The thermal conductivity of various polymer composites reinforced by 3D BN networks, including PDMS (1 − Ref. [215], 1 − Ref. [218], 3 − Ref.[219] , 4 − Ref. [220]), polyamideimide (PAI) (5 − Ref. [213]), EP (6 − Ref. [214], 7 − Ref. [32], 8 − Ref. [217], 9 − Ref. [7a], 10 − Ref. [28], 11 − Ref. [216], 12 − Ref. [31]), polyimide (PI) (13 − Ref. [221], 14 − Ref. [222], 15 −

Ref. [223], 16 – Ref. [224]). NS represents the BNNS-based 3D network, and PL/P represents BN platelets- or particles-based 3D structure.

Comparing with the randomly dispersed BNNSs, the 3D BN network shows much stronger enhancement to the heat transfer in the polymer composite.[32, 217-218] Depending on the fabrication approach, the 3D BN network can also endow the polymer composite with a tailorable anisotropic thermal conductivity.[214] Figure 21e summarizes the thermal conductivity of different polymer composites reinforced by 3D BN networks. It is found that 3D BN network brings a significant enhancement to the thermal transport of the polymer composites compared to the random dispersed BNNS or BN platelets/particles. Based on the Foygel model,[191] the 3D BN network is found to possesses a much smaller filler/filler interfacial resistance compared with the composite with randomly dispersed BNNSs, which leads to a higher thermal conductivity. According to Figure 21e, very few polymer composites have been reported with a $\kappa$ over 5 Wm$^{-1}$K$^{-1}$ with a filler loading within 20vol%. Although some works reported a promising enhancement with a very low volume fraction (e.g., a $\kappa$ of 1.17 Wm$^{-1}$K$^{-1}$ for the 2vol% BNNS/PAI (polyamideimide) composite), it is still unclear whether such high enhancement can be realized in a larger volume fraction. A few works reported a high $\kappa$ over 10 Wm$^{-1}$K$^{-1}$, which however requires a very high filler loading and the 3D structure is based on BN platelets.

Like the CNT-based 3D nanostructured fillers, most of current studies have emphasized the thermal conductivity of polymer composites containing various filler content, under different temperature, or with varying processing parameters/settings. A wide range of factors are shown to affect the heat transfer in the polymer matrix, such as the filler alignment and orientation, filler size and geometry, filler content. These results suggest several key challenges that need to be addressed for engineering application, including how to fabricate and characterize high-quality hBN-based 3D fillers; how to effectively and efficiently control the homogeneous dispersion of the fillers. More importantly, how to minimize the thermal interfacial resistance, especially between the BN filler and the polymer matrix. In addition, to facilitate the engineering applications, it is necessary to investigate how the thermal conductivity of the polymer composites will change under cyclic thermal or mechanical loadings, which still requires great research efforts. An early work show that the thermal conductivity of 3D BNNS/PU (polyurethane) composite maintains a high thermal conductivity after 100 times of bending and twisting tests, which decreases significantly when the tensile strain exceeds ~ 20%.[225]

## 4.3. Ceramic Fillers

In addition to hBN, other high conductivity inorganic ceramics have also been employed to reinforce polymer composites, such as SiC, $Si_3N_4$, $Al_2O_3$, and AlN.[226] They are normally in the form of NW or nanoparticle (NP), and being randomly dispersed in the polymer composites via solution mixing or milling. Due to the strong interfacial thermal resistance, the enhancement on the heat transfer from the random dispersion is limited. Increasing the filler loading can help establish a percolation network and further enhance the heat transfer in the composite, while it will generally degrade the mechanical properties of the composite remarkably. In this regard, recent works have focused on the ceramic fillers based on 3D aligned NWs or networks.

SiC is one of the popular ceramic fillers being used to in the polymer composites. Different methods have been developed to construct 3D SiC networks based on SiC NWs,[227] or Si NWs and graphene NS.[228] The intrinsically different thermal conductivities of the 1D NW (along the axial and lateral directions) brings the opportunities to prepare the polymer composite with anisotropic thermal conductivities.[227] **Figure 22**a compares the thermal conductivity of the EP composite at different filler loadings. Despite different types of SiC fillers, the EP composites show a similar $\kappa$ less than 0.4 Wm⁻¹K⁻¹ in the low filler loading regime (< 2wt%). With the further increasing of the filler loading, the 3D SiC NW network is found to induce a significant enhancement to the thermal conductivity, e.g., a $\kappa$ of about 1.67 Wm⁻¹K⁻¹ at the weight fraction of 5.6%. In comparison, the EP composites with randomly dispersed SiC NWs and NPs show a much smaller $\kappa$, e.g., $\kappa$ is ~ 0.36 Wm⁻¹K⁻¹ at 10 wt% of SiC NP. Such observation is reasonable as the NPs are not able to form effective pathways for the thermal transport compared with that of the NWs. At low filler content, the filler/matrix interface thermal resistance dominantly affect the heat transfer in the polymer composites. Whereas, at higher filler loading (with effective percolation network), the filler/filler interface thermal resistance dominantly affect the heat conduction.[229] Therefore, for 3D networks with the presence of percolation network, the filler/filler interfacial thermal resistance is the main factor that impair effective heat transfer. Based on the Foygel model,[230] the 3D interconnected network possesses a much smaller contact resistance compared with that of the random dispersed NW network.[231] To note that, excessive loading would weaken the heat transfer by creating voids or defects in the polymer matrix. Such structural defects will induce strong phonon scattering and degrade the thermal conductivity in the polymer composites.[232]

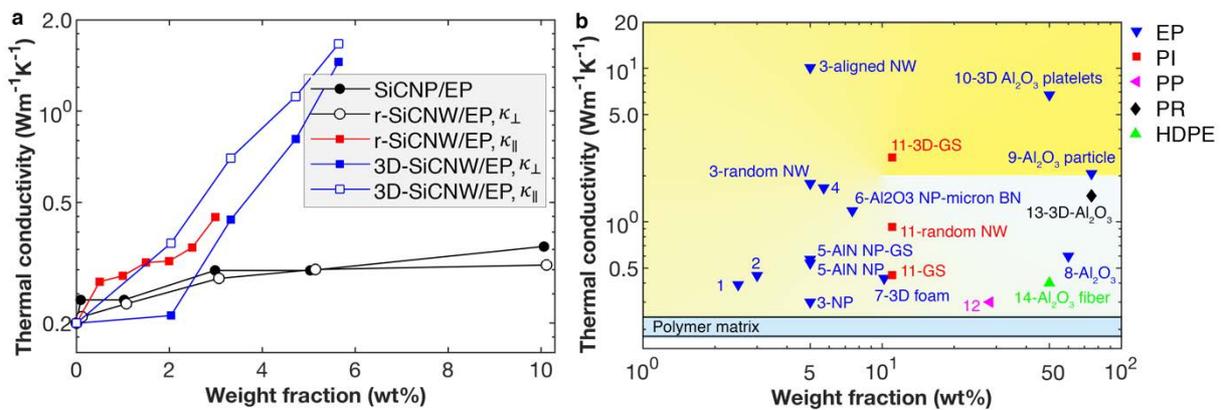

**Figure 22.** The thermal conductivity of polymer composites reinforced by ceramic fillers. (a) The thermal conductivity of the epoxy composite as a function of the weight fraction of SiC

NPs (SiCNP/EP),[232] randomly dispersed SiC NWs (r-SiCNW/EP),[233] and 3D SiC NWs

(3D-SiCNW/EP).[227] The volume fraction is converted to the weight fraction based on a

density of 1.175 g·cm³ and 3.2 g·cm³ for the epoxy matrix and SiC NW, respectively.[228a]

The thermal conductivity of the epoxy matrix is 0.2 Wm⁻¹K⁻¹. $\kappa_\parallel$ and $\kappa_\perp$ represent the in-plane

and out-of-plane thermal conductivity. (b) A summary of the thermal conductivity of the

polymer composited reinforced by ceramic fillers, including EP composite (1 – Ref.[234], 2 –

Ref. [233], 3 – Ref. [232], 4 – Ref. [227], 5 – Ref. [235], 6 – Ref. [236], 7 – Ref.[231], 8 – Ref. [237], 9 –

Ref. [238], 10 – Ref. [239]), PI (9 – Ref. [228b]), Polypropylene (PP, 10 – Ref. [240]), phenolic resin

(PR, 11 – Ref. [241]), and high density polyethylene (HDPE, 12 – Ref. [242]).

Comparing with the carbon-based and hBN-based fillers, the investigation on the

thermal transport in the polymer composites reinforced by other ceramic fillers are still

limited. The filler content is the most frequently discussed influential factor, and only a very

few studies have explored the temperature influence. A recent work shows that the $\kappa$ of

epoxy composites containing 3wt% of SiC NWs increases with temperature.[233] Figure 22b

summaries the thermal conductivity of polymer composites reinforced by different ceramic

nano-fillers. Comparing with the hBN-based filler (in Figure 21b), the SiC, Al₂O₃, and AlN

fillers are still lacking of investigation, and only a few studies have reported the polymer

composites possessing a thermal conductivity larger than 1 Wm⁻¹K⁻¹. Most of the studies are

based on epoxy composites. The highest reported $\kappa$ is 10.10 Wm⁻¹K⁻¹ with 5wt% highly

aligned SiC NWs,[232] which is benefited from the good alignment of the NWs and their high

aspect ratio. Overall, although the ceramic fillers are normally low-cost, to reach a $\kappa$ in the

range of ~ 1 – 5 Wm⁻¹K⁻¹ at room temperature normally requires a high-volume fraction (up

to 70 vol%).[243] Unfortunately, high filler content will not only seriously deteriorate the

mechanical properties of the composites,[244] but also deteriorate their dielectric properties.[204]

## 4.4. Metallic Fillers

Besides the carbon-based and ceramic-based fillers, metallic nanofillers are also appealing candidates to enhance the heat transfer in polymer composites. They are often in forms of NP[245] and NW[244, 246] with a high aspect ratio (AR, length over diameter). As aforementioned, NWs are preferable for TIMs as they can form effective percolation network at a low filler content.[191] The Au[247] or Ag NWs-based fillers are frequently adopted due to their excellent thermal conductivity and resistance to surface oxidation, whereas their industrial applications are limited due to the high cost. On the other hand, Cu NWs-based fillers are attracting increasing attention in recent years majorly because of their much lower cost compared to Au or Ag (while they also have a high thermal conductivity ~ 400 Wm$^{-1}$K$^{-1}$). However, the drawback of Cu NWs is that they suffer from oxidation even in composites, which will result in detrimental influence on the long-term performance of the composites.

Due to the high thermal conductivity of metallic NW filler, the heat transfer in the polymer composites can be remarkably enhanced at a very low filler loading. For instance, with a low volume fraction of 0.9% of Cu NWs (AR = 100 ~ 1000), polyacrylate (PA) composite is reported to achieve a high $\kappa$ of 2.46 Wm$^{-1}$K$^{-1}$.[246b] Together with the high thermal conductivity, metallic NWs also possess a high electrical conductivity, which usually ruins the electrical insulating properties of the polymer composites. For example, the PA composites with 0.9vol% of Cu NWs is measured with an electrical conductivity of 0.04 Scm$^{-1}$, which are no longer electrical insulting.[246b] To overcome this issue, different surface modifications have been proposed for metallic NWs, e.g., grow a nano-coating of SiO$_2$[248] or polydopamine (PDA)[244] on the NW surface. For the Cu NW, the coating can not only be used to resume the electrical insulating feature of the polymer, but also function as surface protector for the NWs from oxidation.

**Figure 23**a shows the dependency of the thermal conductivity of polymer composites on the loading of Cu NW fillers. In general, $\kappa$ increases when the filler loading increases, which is commonly in a nonlinear fashion. The enhancement can be understood from the perspective of the percolation threshold theory.[249] At a low filler content regime, the NWs can barely form a connected network, which greatly hinder the electronic and phonon thermal conduction and thus $\kappa$ exhibits a gradual increasing tendency with the filler content. When an effective percolation network is formed at higher filler loading, the high thermal conductivity of the NWs will significantly enhance the heat transfer, as is seen from the PA composite with long Cu NWs in Figure 23a. Since long NWs are beneficial for the conductive network formation, the composites exhibit a much high $\kappa$ reinforced by long NWs in comparison with the counterpart reinforced by short NWs (under a same filler content).[246b] Experimental results reveal that a proper coating can not only resume a high electrical resistivity, but also enhance the phonon propagation across the NW/matrix interface and promote a further enhance to the heat transfer in the composites.[244, 250] Specifically, the coating layer should have a proper thickness to achieve the highest $\kappa$, and too thick a coating layer will lead to smaller $\kappa$.[244] Different theoretical models have been adopted to interpret the relationship between $\kappa$ and the filler volume fraction, e.g., the Maxwell model,[186] and the Agari model.[251] It is found that the Nan model (that includes interfacial thermal resistances)[185] can provide a reasonable description for the relationship between $\kappa$ and the filler content.

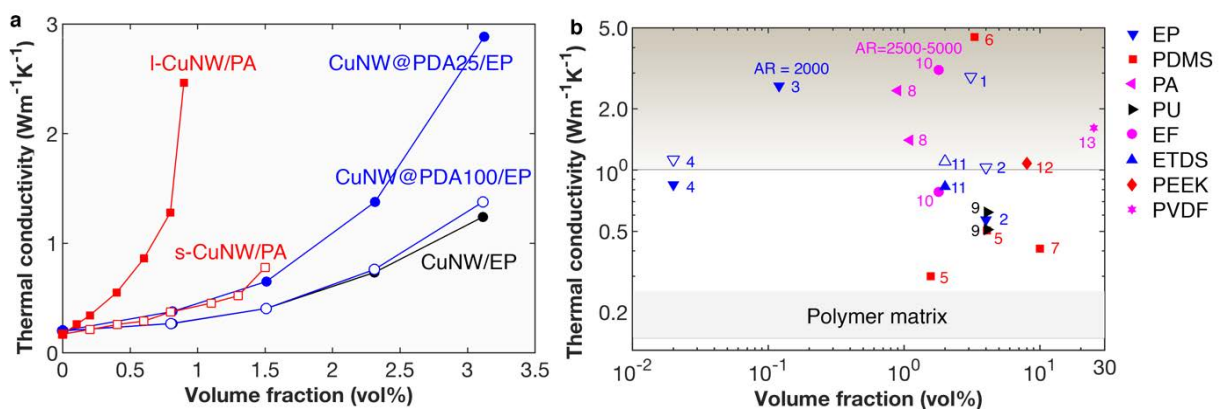

**Figure 23.** Thermal transport in polymer composites with metallic NW fillers. (a) The changing tendency of the thermal conductivity at different Cu NW filler contents for PA composite with short and long Cu NW (denoted as s-CuNW/PA and l-CuNW/PA, respectively),[246b] and epoxy (EP) composite without and with 25 nm and 100 nm PDA coated Cu NW (denoted as CuNW/EP, CuNW@PDA25/EP, and CuNW@PDA100/EP, respectively).[244] (b) A summary of the thermal conductivity for metallic NWs reinforced polymer composites, including EP composite (1 – Ref. [244], 2 – Ref. [250b], 3 – Ref. [252], 4 – Ref. [250a]), PDMS (5 – Ref. [253], 6 – Ref. [247], 7 – Ref. [254]), polyacrylate (PA, 8 – Ref. [246b]), PU (9 – Ref. [253]), EcoFlex (EF, 10 – Ref. [255]), epoxy-terminated dimethylsiloxane (ETDS, 11 – Ref. [248]), polyether ether ketone (PEEK, 12 – Ref. [256]), and polyvinylidene fluoride (PVDF, 13 – Ref. [254]). The open markers represent the NWs with coatings.

Since the thermal conduction is dominant by electrons in the metallic NWs, the thermal conductivity of the polymer composites normally decreases with increasing temperature.[244, 246b, 255] Figure 23b summarizes the thermal conductivity of various polymer composites reinforced by metallic NWs. It is seen that long NW fillers are preferable due to the easy formation of effective conductive networks in the polymer composites.[252] The heat transfer can be further enhanced by reducing the filler/filler resistance. For instance, through microwave radiation treatment, the thermal conductivity of the EcoFlex (a platinum-catalyzed addition-type silicone elastomer) composite with 1.8vol% Cu NWs can be significantly improved from 0.78 $Wm^{-1}K^{-1}$ to 3.1 $Wm^{-1}K^{-1}$.[255] It is worthy to mention that besides NWs, other metallic multiscale fillers have also been used to enhance the thermal transport in polymer composites. For instance, for the EP composite with 3D microscale Ag flakes (around 36.7% vol%), an addition of around 2.3 vol% of multi-wall CNTs (decorated with Ag NPs) can results in a significant enhance to the thermal conductivity from 1.64 $Wm^{-1}K^{-1}$ to 160 Wm$^{-1}$K$^{-1}$.[257] Such remarkable enhancement is supposed as resulted from the efficient phonon transport pathways between Ag flakes that are bridged by the CNT (with the aid of Ag NPs).

Overall, current efforts have been mainly devoted to identify the relationship between the thermal conductivity of the polymer composites and the filler loading. Some works also discussed the temperature influence on the heat transfer. Comparing with the carbon-based and hBN-based fillers, the metallic NW fillers are normally randomly dispersed in the polymer matrix. Although the heat transfer in the polymer matrix with pre-prepared 3D metallic networks are still unexplored, the challenges regarding the filler preparation, filler characterization, and dispersion are still expected for the metallic nanofillers. Furthermore, the combination of electron and phonon transport is expected to bring more uncertainties to the bottlenecks that hinder the effective heat transfer, i.e., the interfacial thermal resistance between the metallic filler and the polymer matrix.

## 5. Conclusion and Perspectives

Diverse 3D nanostructures have been fabricated recently to meet specific thermal management purposes, ranging from nanoarchitectures to nanostructured fillers for polymers and metals. Here, we have outlined the recent progress on the thermal conductivity of those nanostructures with a focus on experimental results.

### 5.1. Low thermal conductivity

To pursue low thermal conductivity for thermal isolation application, different types of 3D nanoarchitectures with a high porosity have been prepared based on low-dimensional nanostructures. Recent years have witnessed a growing interest in preparing colloidal assemblies to achieve low thermal conductivity, which can be fabricated from single or multiple colloidal particles (mostly based on polymers). Studies have identified several influential factors that affect the thermal transport of the colloidal assemblies, including particle size, composition, and arrangement or packing. The intrinsic phase change characteristic of the polymer can endow colloidal assemblies with a unique temperature dependence, which enables emerging heat management applications in phase change materials. More sophisticated core-shell particles have also been prepared to fabricate the colloidal assemblies with improved mechanical performance while maintaining a low heat transfer. The majority of current studies have focused on spherical particles, it is still unknown how the particle shape would impact the thermal transport performance. In considering the actual usage, significant efforts are still desired to correlate the thermal transport properties of the colloidal assemblies with other application-specified properties (e.g., mechanical properties or transparency) while tailoring their structures.

In addition to colloidal crystals, many other 3D nanostructures have also been prepared for thermal isolation, such as nanocellular foams and nanolattice. The polymer-based nanocellular foams have been extensively explored, however the investigation on the thermal transport properties of the nanolattice, is still rare. Given the rapidly developing additive manufacturing techniques, it is promising to design and fabricate nanolattice with desired thermal transport properties together with excellent mechanical performance for thermal isolation applications.

In terms of the thermoelectric devices, researchers have considered the array assembles of nanowires, the inverse opals structures and holey nanostructures. The thermal transport properties of the inverse opals structures are still lacking investigation. While, considering the large variety of materials (e.g., metals and semiconductors) and structures from the sacrificial opal template, there are great potentials to design novel heat management materials with suppressed heat transfer for TE devices, or efficient heat transfer for thermal storage and heat exchangers. The holey 3D Si nanostructures or Si phononic crystals have been shown with a good control over the phonon propagation, which have attracted increasing attention in recent years. The enhanced phonon boundary and surface scattering in the porous structure are proposed to suppress the thermal conductivity. Relationships between the geometrical factors and the thermal conductivity of the holey nanostructures have been extensively investigated, such as the diameter and arrangement. However, although it is widely accepted that the nanoscopic holes in the phononic crystals suppress the thermal conductivity, the underlying mechanism is still under argument. Besides, using the phonon coherence to tune the thermal transport in phononic crystals is still limited at low temperature. However, it is believed that with the advancing of nanofabrication technology, the working temperature range of phononic crystals can be further broadened like the way that wave optics revolutionized the manipulations of light.

To summarize, the porous or hollow structures are main players for thermal insulation applications, which are commonly prepared from materials with inherently low thermal conductivity such as polymers and silica. The heat transfer in the hollow structures contains the contribution from radiation, convection, gas phase conduction and solid phase conduction, among which the gas phase conduction and solid phase conduction are the main contributors. Therefore, suppress the contributions from the gas phase and solid phase are the main avenues to obtain a low thermal conductivity. It has been shown that the gas phase conduction can be significantly suppressed when the pore sizes in the hollow structures are on the order of mean free path of gas molecules, in which circumstance the gas changes into Knudsen flow. Such concept has been successfully implemented and proved in the 3D polymer-based foams or nanocellular foams. On the other hand, phonons are the main heat carrier for solid phase conduction, and the factors that influence the phonon transport are all useful means to engineer the thermal conductivity of the porous structures. For instance, in the colloidal crystals/assemblies, the interface phonon scattering, the distorted and elongated heat transfer pathways, and the porosity are all effective ways to restrain heat transfer in the structure.

There are generally two strategies to attain high efficiency in TE devices, i.e., phonon engineering to reduce the lattice thermal conductivity and the band engineering to boost the electrical power factor. Nanostructuring has been shown as an effect avenue to suppress phonon transport by enhancing phonon scattering (including Umklapp phonon-phonon scattering, phonon-impurity scattering, phonon-boundary scattering) or introducing phonon confinement effect. Recent years, great interests have been attracted on the manipulation of the so-called wave-like coherent phonon transport, which can be realized through the presence of a secondary periodicity in the nanostructure. The phonon coherence has already been demonstrated to result in remarkably reduced thermal conductivity in phononic crystals, which opens up a new and promising avenue to construct high-performance TE devices.[99]

## 5.2. High thermal conductivity

High thermal conductivity is desirable for the thermal interface materials (TIMs) and phase change materials (for thermal energy storage), and recent years have witnessed a significant research focus on TIMs. One promising TIM candidate is the CNT array, which have been shown with excellent thermal conductivity. The emerging challenge is how to minimize the thermal contact resistance between CNT arrays and the mating surface, and how to fabricate well-aligned high-quality CNT arrays.

Plenty of in silico studies have been carried out on 3D nanoarchitectures (mainly constructed from CNT or graphene), such as pillared structures and honeycomb, while the experimental synthetization, characterization and measurement are still very challenging. Several works reported the thermal conductivity for some of the hierarchical structures (containing graphene/graphite and CNT), which however exhibit a large variance. Given the vital role of the influence of the structure on the energy carrier, it is hard to compare the results between different structures. Compared with the CNT arrays, these 3D nanoarchitectures are still far away from real applications.

Another extensively explored TIM candidate is the polymer composites, and various types of highly conductive nanostructures have been utilized as fillers to enhance their thermal conductivity, such as nanowire and 2D nanosheet (ranging from carbon, ceramic, to metallic nanomaterials). Compared with the intrinsically low thermal conductivity of the polymer ($\sim 0.2$ $Wm^{-1}K^{-1}$), remarkable enhancement has been reported. However, the thermal conductivity for most of the reported polymer composites are still well below 5 $Wm^{-1}K^{-1}$. To attain higher thermal conductivity, the filler content needs to be high, which inevitably brings negative impacts on the mechanical performance or the electrical insulation characteristic of the composites. Specifically, compared with the intrinsic thermal conductivity of the nanofillers, there is still a significant gap between the effective thermal conductivity of the composite and expectations (e.g., $\sim 3000$ for $Wm^{-1}K^{-1}$ CNTs). There is a tendency to use multiple (two or more) fillers to reach a high thermal conductivity at a relatively lower filler content utilizing the synergistic effect of the fillers. Again however, the improvement is still far below expectation.

A wide range of factors have been shown to significantly affect the heat transfer in the polymer composites, such as the filler content, filler/matrix interface, filler/filler interface, filler geometry and size, filler alignment and orientation, and the fabrication/processing techniques. It is widely accepted that a percolation network of fillers is desirable in the polymer matrix to achieve effective heat transfer. The dispersion of filler in a polymer matrix plays a crucial role for the heat transfer in the polymer. Similar as the CNT arrays, the bottleneck to obtain highly conductive polymer composites is to minimize the thermal interfacial resistance within the matrix.

There is also increasing interest in adding nanostructured fillers to metal matrix in the past decade, in the purpose of reducing the coefficient of thermal expansion and enhancing their thermal conductivity. According to present studies, the coefficient of thermal expansion can be reduced by the nanofillers, which however commonly bring negative influence to the heat transfer in metal matrix. Currently, it is challenge to prepare metal matrix composites with desired nanofillers and to identify the bonds at the interface. As such great efforts are still expected in the future to explore the influential factors on the heat transfer in the metal matrix, such as temperature, filler shape and geometry, and filler dispersion or network.

Overall, there is an increasingly growing demand for high thermal conductivity materials due to the rapidly advancing electronic industry, especially the thermal interface materials for electronic packing. Majority works have been focused on CNT-based or polymer composite-based TIMs, where the phonon is the dominant heat carrier. Therefore, a key feature for TIMs is having efficient heat transfer channel to spread undesired heat. CNT arrays are good example with excellent thermal conductivity. For polymer composites, adding highly thermal conductive fillers has been shown as an effective way to promote the heat transfer. However, it is urgently needed to find out how to minimize the thermal interfacial resistance that introduces strong phonon scattering.

To conclude, the use of nanostructuring of materials is an effective strategy to modify the thermal conductivity of materials (either by phonon or electron transport). Studies have shown that a broad range of factors, such as structure (or morphology), geometrical size, and structural defects, will affect the heat carrier. To facilitate engineering applications, it is vital to explore and develop cost-effective and efficient approaches to prepare high-quality nanostructures. In addition, it is still an open question whether these nanostructures/nanomaterials can retain their efficient heat transfer capability under repetitive thermal or mechanical loadings during service, which is an important area for future research.

**Acknowledgements**

H.Z., J.B. and Y.G. would like to acknowledge the supports by the ARC Discovery Project (DP170102861). Y.C. would like to acknowledge the financial support by Key projects of Shaanxi Natural Science Foundation (2019JZ-27), Shaanxi Natural Science Basic Research Program-Shaanxi Coal (2019JLM-47) and Fundamental Research Funds for the Central Universities CHD (300102319304).

WILEY-VCH

**Author Bio-data:**

**Haifei Zhan** is currently a Lecturer of Mechanical Engineering at the Queensland University of Technology. He received his Ph.D. degree in mechanical engineering from there in 2013, and worked as Postdoc Research Fellow until 2016. In 2017, he worked as a lecturer in Western Sydney University and moved back to Queensland University of Technology in 2018. His research focuses on nanoscale thermal transport properties, and mechanical properties of low-dimensional nanomaterials, nanocomposites, and nanofibers.

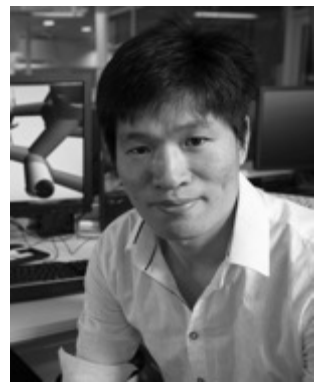

**Yongnan Chen** is currently a Professor of Material Engineering at Chang'an University. He received his Ph.D. degree in material science from Xi'an Jiaotong University in 2009, and then moved to Chang'an University since 2010. His research focuses on the synthetization, characterization and commercialization of high-performance metals/alloys and metallic composites, and wear and corrosion protection.

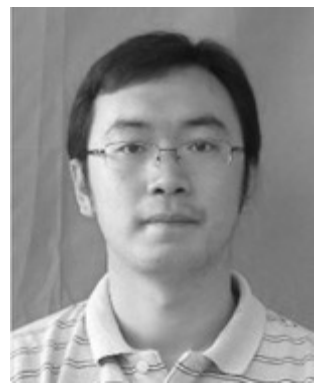

**Yuantong Gu** is currently a Professor of Mechanical Engineering at the Queensland University of Technology. He received his Ph.D. degree in mechanical engineering from National University of Singapore in 2003 and worked as a Research Fellow in the University of California (Irvine) in 2004 and The University of Sydney from 2005. In 2007, he moved to Queensland University of

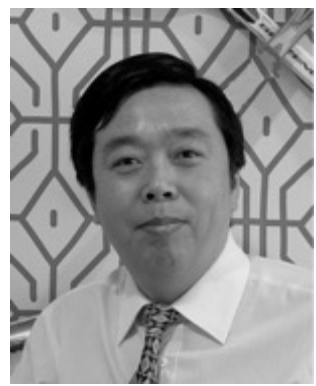

Technology. His research focuses on the computational mechanics, nanomechanics, and biomechanics.

**WILEY-VCH**

**Table of Contents**

**Keyword**: Thermal conductivity, nanostructures, carbon nanotube, nanocomposite, metal matrix composites

H.F. Zhan, Y.H. Nie, Y.N. Chen*, J. M. Bell, Y.T. Gu*

**Title**: Thermal Transport of 3D Nanostructures

**ToC figure:**

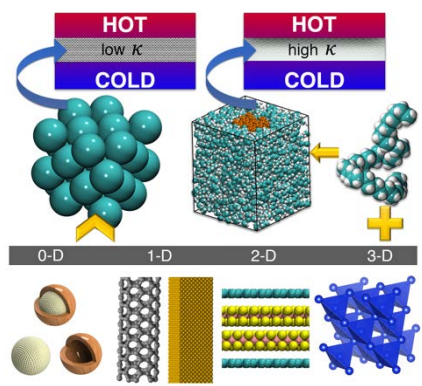

**ToC entry:** Diverse 3D nanostructures have been fabricated to meet specific thermal management purposes, including nanoarchitectures (like colloidal assemblies, carbon nanotube arrays), and nanocomposites from polymer matrix or metal matrix. It is shown that the use of nanostructuring of materials is an effective strategy to modify the thermal conductivity of materials.